\documentclass[aps,twocolumn,prb,amsmath,amssymb]{revtex4-1}
\usepackage[varg]{txfonts}
\usepackage{color}
\usepackage[colorlinks]{hyperref}
\usepackage{amsmath}

\usepackage{graphicx}
\usepackage{bm}

\def\XXint#1#2#3{{\setbox0=\hbox{$#1{#2#3}{\int}$}
    \vcenter{\hbox{$#2#3$}}\kern-.5\wd0}}

\def\be{\begin{equation}}
\def\ee{\end{equation}}

\def\bi{\begin{itemize}}
    \def\ei{\end{itemize}}
\def\bn{\begin{enumerate}}
    \def\en{\end{enumerate}}
\def\bea{\begin{eqnarray}}
\def\eea{\end{eqnarray}}
\newcommand{\bpm}{\begin{pmatrix}}
    \newcommand{\epm}{\end{pmatrix}}

\def\ba{\begin{array}}
    \def\ea{\end{array}}
\def\bd{\begin{displaymath}}
\def\ed{\end{displaymath}}

\renewcommand{\imath}{\hspace{1pt}\mathrm{i}\hspace{1pt}}

\begin{document}

\title{Hybrid topological magnon-phonon modes in ferromagnetic honeycomb and kagome lattices}

\author{Bahman Sheikhi}
\affiliation{Department of Physics, Sharif University of Technology, Tehran 14588-89694, Iran}

\author{Mehdi Kargarian}
\email{kargarian@physics.sharif.edu}
\affiliation{Department of Physics, Sharif University of Technology, Tehran 14588-89694, Iran}

\author{Abdollah Langari}
\affiliation{Department of Physics, Sharif University of Technology, Tehran 14588-89694, Iran}

\begin{abstract}
Magnons and phonons are two fundamental neutral excitations of magnetically ordered materials which can significantly dominate the low-energy thermal properties. In this work we study the interplay of magnons and phonons in honeycomb and Kagome lattices. When the mirror reflection with respect to the magnetic ordering direction is broken, the symmetry-allowed in-plane Dzyaloshinskii-Moriya (DM) interaction will couple the magnons to the phonons and the magnon-polaron states are formed. Besides, both lattice structures also allow for an out-of-plane DM interaction rendering the uncoupled magnons to be topological. Our aim is to study the interplay of such topological magnons with phonons. We show that the hybridization between magnons and phonons can significantly redistribute the Berry curvature among the bands. Especially, we found that the topological magnon band becomes trivial while the hybridized states at lower energy acquire Berry curvature strongly peaked near the avoided crossings. As such the thermal Hall conductivity of topological magnons shows significant changes due to coupling to the phonons.               

\end{abstract}

\maketitle

\section{Introduction}\label{Introduction}
The notion of band topology has become ubiquitous in condensed matter systems such as band insulators\cite{Hasan:RMP2010, Hasan:Annals2011} and superconductors\cite{Qi:RMP2011} in two- and three-dimensional spaces, Weyl and Dirac semimetals\cite{Armitage:RMP2018,Burkov:Annals2018}, photons\cite{Raghu:PRA2008, Hafezi:NatPhys2011, Khanikaev:NatMat2013}, magnons\cite{Katsura:PRL2010, Shindou:PRB2013}, phonons\cite{Strohm:PRL2005,Zhang:PRL2010,Yang:PRL2015,Susstrunk:pnas2016}, plasmons\cite{MagnetoPlasmons1,MagnetoPlasmons2}, and even in classical systems\cite{Kane:NatPhys2014, Paulose:NatPhys2015,Huber:NatPhy2016}. In another frontier, attempts put forward to induce topological phases by proper combinations of otherwise trivial states. Famous examples include the Floquet topological insulators, where a periodically driven electronic system by light becomes topologically nontrivial\cite{Galitski:NatPhys2011}, and topological polaritonic states
emerge from the interaction of single photons with excitons\cite{TopPolaritons1}. Recently, in a work by one of the authors, it is shown that a proper hybridization of spin and plasma waves gives rise to topological collective modes\cite{Efimkin:arXiv2020}.      

Magnons and phonons are two fundamental collective excitations in solids. Magnons are quanta of collective spin waves in magnetically ordered solids, and phonons describe the elastic modes of a solid. The topological properties of magnons and phonons have been extensively studied. Both excitations have neutral charge and, usually the thermal Hall conductivity measurements are used to probe the possible nontrivial dynamics of thermal carriers. The experimental observation of magnon mediated thermal Hall response in ordered magnet Lu$_2$V$_2$O$_7$ \cite{Onose:Science2010}, planar kagome magnet Cu(1,3-benzenedicarboxylate) \cite{Ong:PRL2015}, and in metallic ferromagnetic kagome YMn$_6$Sn$_6$\cite{Zhang:prb2020} can be understood using the notion of topological magnons\cite{Katsura:PRL2010}. The topological magnons have been studied in three dimensional structures\cite{Kondo:PRB2019, Hwang:PRL2020}, honeycomb\cite{Moulsdale:PRB2019} and kagome\cite{Owerre:PRB2017,Laurell:PRB2018} lattices, and in pyrochlore thin films\cite{Laurell:PRL2017}. The phonon Hall effect has been observed in the paramagnetic dielectric Tb$_3$Ga$_5$O$_{12}$\cite{Strohm:PRL2005,Inyushkin:JETP2007}, and the phenomenon is associated to magnetic ions coupled to lattice vibrations \cite{Sheng:PRL2006,Kagan:PRL2008, Mori:PRL2014,Saito:PRL2019}. Also, the anomalous thermal Hall response in the frustrated magnet Tb$_2$Ti$_2$O$_7$\cite{Hirschberger:Science2015} is attributed to phonons \cite{Hirokane:PRB2019}.  

On the other hand, and besides the applications in magnon  spintronics\cite{Bozhko:AIP2020}, the interaction between magnons and phonons may also lead to thermal Hall effect. In Ref.~[\onlinecite{Takahashi:PRL2016}], it is shown that the interaction between magnons and acoustic phonons induces the Berry curvature which can modify the dynamics of the wave packets. Using a spin-phonon model on the square lattice, the in-plane Dzyaloshinskii-Moriya (DM) interaction, resulting from the mirror symmetry breaking, hybridizes the magnons and phonons. The generated Berry curvature at the crossings of magnon and phonon energy bands induces the thermal Hall conductivity \cite{Zhang:PRL2019,Go:PRL2019,Park:Nano2020}, while in the absence of interaction neither magnons nor phonons carry thermal Hall response. 

In this work, we study spin systems coupled to phonons in the honeycomb and kagome lattices. The spin models on these lattices, when endowed with proper DM interactions, allow for realization of topological magnons. While in previous works both bosonic modes are trivial, we designate the spin model to yield topological magnons, yet, the phonons are trivial in the absence of interaction between them. We ask the following questions: (i) how does the nontrivial band topology of magnons influence the phonons? (ii) how does the Berry curvature redistribute among the hybridized energy bands? and (iii) how does the latter hybridization, resulting in magnon-polaron excitations, reflect in the thermal Hall conductivity of magnons due to coupling to trivial phonons?. For comparison, on the honeycomb lattice we also compute the thermal Hall conductivity of hybridized modes otherwise being trivial when decoupled. For the case of the honeycomb lattice, the coupling of topological magnons to phonons has also been studied in Ref.~[\onlinecite{Thingstad:PRL2019}], where the spin is coupled to out-of-plane displacements\cite{Kittel1,Kittel2}. In our model systems, however, we couple the spin fluctuations to in-plane displacements through the DM interaction which could be more relevant in two-dimensional and layered materials. The observation of giant thermal magnetoconductivity in the layered compound CrCl$_3$, an insulating magnet with underlying honeycomb lattice, indicates that the phonons and their scattering from the magnons play an important role\cite{Pocs:PRR2020}. In cubic antiferromagnetic compound Cu$_3$TeO$_6$,  whose magnon excitations are shown to be topological\cite{Li:PRL2017,Bao:NatComm2018}, the inelastic neutron scattering provides compelling evidence of magnon-phonon coupling\cite{Bao:PRB2020}. The same measurements on hexagonal multiferroic YMnO$_3$ reveal a gap opening below the ordering temperature rendering low-energy excitations with strong coupling between magnons and phonons\cite{Petit:PRL2007}. The hybridization of magnons with phonons in the topological semimetal Mn$_3$Ge, a breathing kagome antiferromagnet, has been employed to understand the neutron scattering measurements\cite{Dasgupta:PRB2020, Dasgupta2:PRB2020}. Also, the optically excited magnon-phonon hybrid excitations has also been reported in nanograting galfenol Fe$_{0.81}$Ga$_{0.19}$ \cite{Godejohann:PRB2020}.

\begin{figure}[t]
   \center
    \includegraphics[width=\linewidth]{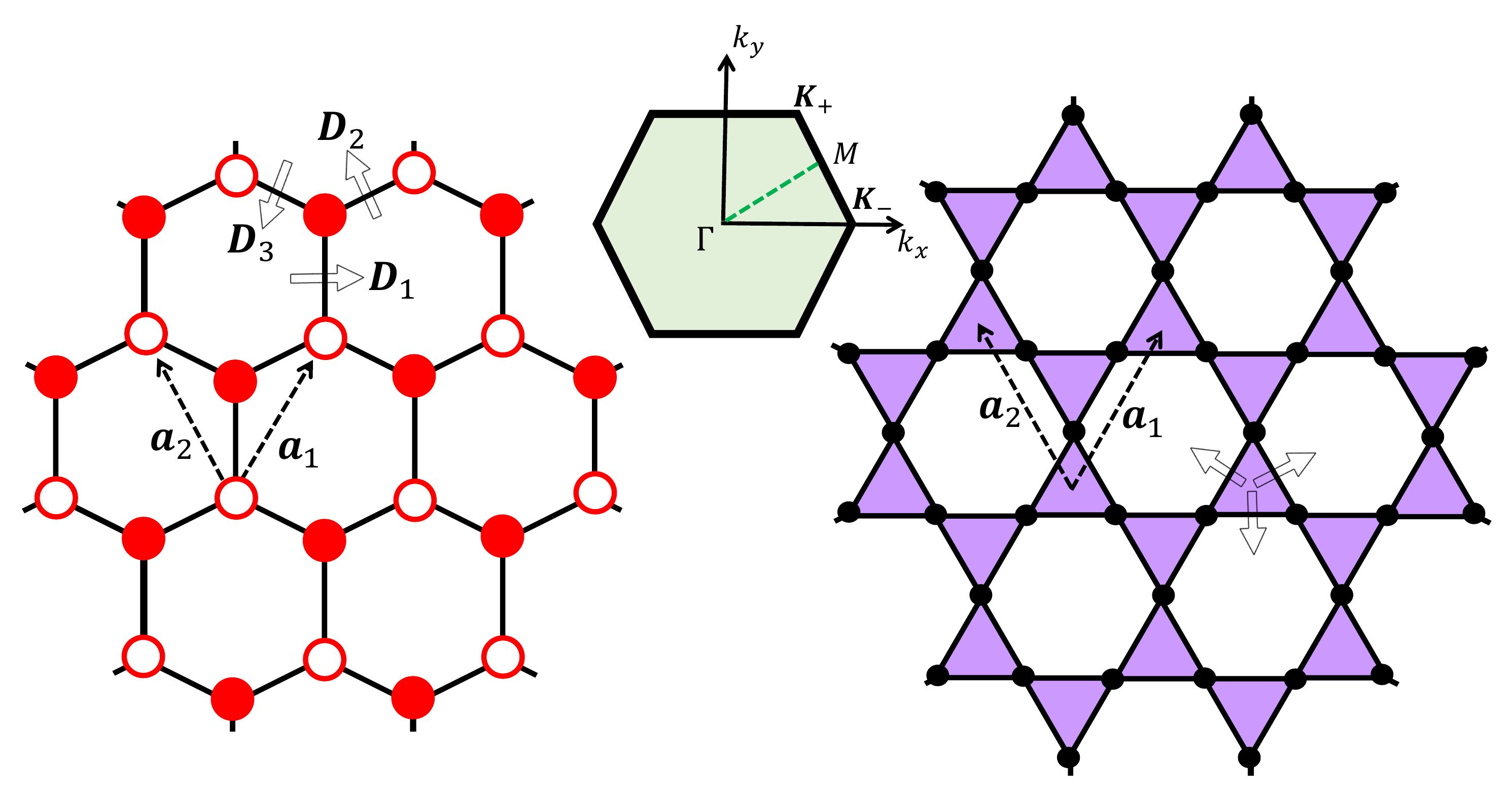}
    \caption{A schematic of honeycomb (left) and regular kagome (right) lattices studied in this work. Primitive lattice vectors $\mathbf{a}_1$ and $\mathbf{a}_2$ are shown by dashed arrows. Double arrows indicate the directions of the in-plane  Dzyaloshinskii Moriya vector $\mathbf{D}_{\parallel}$ on nearest neighbor bonds. In the middle top, the Brillouin zone is shown with high symmetry points indicated.} \label{lattice}
\end{figure}

Therefore, the above observations call for a deep investigation of magnon-phonon couplings and their impacts on physical properties. Especially, we aim at studying the topological contents of magnon-polaron excitations on the honeycomb and kagome lattices. After introducing the magnon-phonon model, we elaborate on the questions posed above. In particular, we derive an effective model describing the hybridization between magnons and phonons and the enhancement of Berry curvature in the vicinity of the avoided crossings. We show that the Berry curvature of high-energy magnons are redistributed to low-energy hybrid bands. As such the intrinsically magnon-mediated thermal Hall conductivity substantially changes upon coupling to phonons. For the kagome lattice, in addition to the regular lattice shown in Fig.~\ref{lattice}, we also consider a distorted lattice obtained by solid twistings of the unit cells (see Fig.~\ref{tKagome}(a)) and study the coupling to the magnons. The twisting results from the particular structure of the short-range inter-ion potential energy on the kagome lattice.     

The paper is organized as follows. In Sec.\ref{ModelH} we derive the magnon-phonon Hamiltonian on the honeycomb lattice and study the magnetoelastic spectrum,  the Berry curvature of energy bands, and the thermal Hall response in Sec.\ref{BerryHoney}. The case of kagome lattice is studied in Sec.\ref{Sec:Kagome}, and we conclude in Sec.\ref{conclusion}. Some details of derivation of Hamiltonians are relegated to appendices.

\section{Hybrid magnon-phonon model: the honeycomb lattice \label{ModelH}}
We begin our discussion of hybrid modes by considering the following Hamiltonian,
\begin{align}\label{Htotal}
H=H_{m}+H_{ph}+H_{c}.
\end{align}

In this model $H_{m}$ describes the magnetic Hamiltonian,  $H_{ph}$ gives the phonon dynamics and vibrational modes of the systems, and the last term $H_{c}$ accounts for the coupling between magnetic excitations and phonons. In subsections below  we will describe each Hamiltonian, separately.

\subsection{Magnetic Hamiltonian\label{Sec:magnetic}}
The magnetic Hamiltonian consists of magnetic interactions between localized magnetic ions residing on the vertices of the honeycomb lattice shown in Fig.~\ref{lattice}:

\begin{align}\label{Hm}
H_{m}=-\sum_{i,j}J_{ij}\mathbf{S}_{i}\cdot\mathbf{S}_{j}+\sum_{<<i,j>>}\mathbf{D}_{ij}\cdot(\mathbf{S}_{i}\times\mathbf{S}_j)
\end{align}
where the first sum runs over nearest and next-nearest neighbor sites with exchange interactions $J$ and $J'$, respectively. We assume that $J, J'>0$, implying a colinear ferromagnetic classical ground state below the transition temperature. The second term between next-nearest neighbours is the antisymmetric Dzyaloshinskii-Moriya (DM) interaction specified as $\mathbf{D}_{ij}=\nu_{ij}D\hat{z}$ with $\nu_{ij}=\pm$, depending on right (+) or left (-) turning of site $j$ w.r.t site $i$. 

The low-energy magnetic excitations, the magnons, are readily described using the Holstein-Primakoff transformation of spins to bosons,
\begin{align}
S_{i}^{+}=\sqrt{2S-b^{\dagger}_{i}b_i}~b_i,\;\; S_{i}^{-}=(S_{i}^{+})^{\dagger},\;\; S_i^{z}=S-b^{\dagger}_{i}b_i,
\end{align} 
where $S_{i}^{\pm}$ are raising and lowering spin operators, and $b_{i}$'s are the bosonic operators $[b_i, b^{\dagger}_{j}]=\delta_{ij}$. Within the linear spin-wave approximation, $S_{i}^{+}\approx \sqrt{2S}~b_i$ and $S_{i}^{-}\approx \sqrt{2S}~b^{\dagger}_i$. The magnon description reads as\cite{Owerre_2016},
\begin{align}\label{Hmag}
H_m=-J_n\sum_{<i,j>}b^{\dagger}_{i}b_j-J_s\sum_{<<i,j>>} e^{-i\phi_{ij}}b^{\dagger}_{i}b_j+h.c.,
\end{align}
where $J_n=JS$, $J_s=S\sqrt{J'^2+D^2}$, and $\phi_{ij}=\nu_{ij}\tan^{-1}(D/J')$. Without loss of generality and for sake of simplicity we assume $J'\rightarrow0$ yielding $\phi_{ij}=\pi\nu_{ij}/2$. In momentum space the magnon Hamiltonian becomes,
\begin{align}\label{HmK}
H_m=\sum_{\mathbf{k}} \mathbf{b}^{\dagger}_{\mathbf{k}} H_{m}(\mathbf{k}) \mathbf{b}_{\mathbf{k}},
\end{align}  
 where $\mathbf{b}_{\mathbf{k}}=(b_{\mathbf{k}A}, b_{\mathbf{k}B})^{T}$ is the vector of boson operators acting on sublattices $A$ and $B$ of the honeycomb lattice. The magnon energy dispersion is given by eigenvalues of the Bloch Hamiltonian $H_{m}(\mathbf{k})=h_{0}(\mathbf{k})\mathbf{1} +\mathbf{h}(\mathbf{k})\cdot\boldsymbol{\sigma}$ with $\boldsymbol{\sigma}$ as vector of Pauli matrices, and 
\begin{align}
&h_x(\mathbf{k})+ih_y(\mathbf{k})=-J_n f(\mathbf{k}),~~f(\mathbf{k})=1+e^{-i\mathbf{k}\cdot\mathbf{a}_{1}}+e^{-i\mathbf{k}\cdot\mathbf{a}_{2}},\\
&h_z(\mathbf{k})=2J_s\left[\sin\mathbf{k}\cdot\mathbf{a}_{1}- \sin\mathbf{k}\cdot\mathbf{a}_{2} -\sin\mathbf{k}\cdot(\mathbf{a}_{1}-\mathbf{a}_2) \right],\\
&h_{0}(\mathbf{k})=3J_n.
\end{align}

Note that the classical ferromagnetic ordering has to be stablized by adding either an onsite anisotropic interaction\cite{Zhang:PRL2019} or a small magnetic field to the magnetic Hamiltonian. In the latter case the only change in the bove expressions is $h_{0}(\mathbf{k})=3J_n+BS$, where $B$ is the magnetic field. Such a small field however does not affect the results significantly and we drop it out for simplicity. The magnon band structure has been extensively studied in the literature. In the absence of DM interaction the magnon energy bands cross each other at the corner of BZ, forming Dirac dispersions, like the electron energy bands in graphene, where $f(\mbox{K}_{\pm})=0$. The DM interaction adds phase winding to magnons via $e^{-i\phi_{ij}}$ in the nearest-neighbor hoppings. This is a  bosonic analogue of the famous Haldane model\cite{Owerre_2016,Kim:PRL2016}. A topological gap is opened at the Dirac nodes and the energy bands are characterized by integer Chern numbers $c=\pm1$.

\subsection{Phonon Hamiltonian}
For a magnetic insulator, besides magnons described above, the vibrational modes also contribute to the low-energy properties of the system. The first quantized Hamiltonian describing the ionic motions is, 

\begin{align}\label{Hph}
H_{ph}=\sum_{i}\frac{\mathbf{p}_{i}^2}{2M}+\sum_{i,j}V(\mathbf{R}_{i,j}),
\end{align}
where the first term is the kinetic energy with $M$ as the ion mass, and the second term accounts for the inter-ion potential energy. Here $\mathbf{R}_{i,j}=\mathbf{R}_{i}-\mathbf{R}_{j}$ is a spatial vector connecting ions. To describe the phonons we consider slight deviations from equilibrium positions, i. e., $\mathbf{R}_{i}=\mathbf{R}_{i}^{0}+\mathbf{u}_i$, where $\mathbf{R}_{i}^{0}$ denotes the equilibrium positions of ions. Up to first nonzero terms in the expansion of potential, we obtain,

\begin{align}\label{V_ion}
V(\mathbf{R}_{ij}^{0}+\mathbf{u}_i-\mathbf{u}_j)\approx V(\mathbf{R}_{ij}^{0})+\frac{1}{2}\sum_{\alpha\beta}u_{ij}^{\alpha}\frac{\partial^2V }{\partial u^{\alpha}_{i}\partial u^{\beta}_{j}}u_{ij}^{\beta},
\end{align}
where $\alpha,\beta=x,y$ and $u_{ij}^{\alpha}=u_{i}^{\alpha}-u_{j}^{\alpha}$. Note that we ignored the out-of-plane vibrations as they constitute energy modes higher than the in-plane modes. For our purposes in this work, we restrict the potential energy in \eqref{V_ion} to only first and second neighbors. Moreover, we consider the linear deviations along the bonds. Hence, we get,

\begin{align}
H_{ph}=\sum_{i}\frac{\mathbf{p}_{i}^2}{2M}+\frac{1}{2}M\sum_{i,j}\Omega_{ij}^2 \left[(\mathbf{u}_i-\mathbf{u}_j)\cdot\hat{R}_{ij}^{0} \right]^2,
\end{align}
where $\Omega_{<ij>}=\Omega$ and $\Omega_{<<ij>>}=\Omega'$ are the bond vibrational frequencies of the first and second neighbors, respectively. Here, $\hat{R}_{ij}^{0}=\mathbf{R}_{ij}^{0}/||\mathbf{R}_{ij}^{0}||$ is the unit vector. 

It is instructive to make the momentum and position operators dimensionless by defining $\tilde{\mathbf{p}}=\sqrt{1/M\Omega\hbar}\mathbf{p}$ and $\tilde{\mathbf{u}}=\sqrt{M\Omega/\hbar}\mathbf{u}$ with the commutation relation as $[\tilde{u}^{\alpha}, \tilde{p}^{\beta}]=i\delta_{\alpha\beta}$. With this change of variables, the phonon Hamiltonian becomes 

\begin{align}\label{HphR}
H_{ph}=\frac{1}{2}\hbar\Omega\left\{\sum_{i}\tilde{\mathbf{p}}_i^2+\sum_{ij}\xi_{ij}\left[(\tilde{\mathbf{u}}_i-\tilde{\mathbf{u}}_j)\cdot\hat{R}_{ij}^{0} \right]^2 \right\},
\end{align} 
where $\xi_{<ij>}=1$ and $\xi_{<<ij>>}=\xi'=\Omega'^2/\Omega^2$. Fourier transformed to momentum space, the phonon Hamiltonian reads as,

\begin{align}\label{Hph}
H_{ph}=\frac{1}{2}\sum_{\mathbf{k}}\phi^{t}_{-\mathbf{k}}\tilde{H}_{ph}(\mathbf{k})\phi_{\mathbf{k}},
\end{align}  
where $\phi_{\mathbf{k}}=(u^{x}_{\mathbf{k}A}, u^{y}_{\mathbf{k}A}, u^{x}_{\mathbf{k}B}, u^{y}_{\mathbf{k}B}, p^{x}_{-\mathbf{k}A}, p^{y}_{-\mathbf{k}A}, p^{x}_{-\mathbf{k}B}, p^{y}_{-\mathbf{k}B})^t$. Note that, hereafter with the abuse of notation we drop the tilde from the dimensionless position and momentum variables. Then, $\tilde{H}(\mathbf{k})$ reads as,

\begin{align}
\tilde{H}_{ph}(\mathbf{k})=\hbar\Omega  
 \begin{pmatrix}
  V(\mathbf{k}) & \mathbf{0}_{4\times4} \\
  \mathbf{0}_{4\times4} & \mathbf{1}_{4\times4}  
 \end{pmatrix},
\end{align}      
with $\mathbf{0}_{4\times4}$ and $\mathbf{1}_{4\times4}$ as zero and identity matrices. The matrix $V(\mathbf{k})=[V_{nn}(\mathbf{k})+\xi' V_{nnn}(\mathbf{k})]/2$ contains the first, $V_{nn}$, and second, $V_{nnn}$, neighbor potential terms between ions. The first neighbor terms are,
\begin{align}
&V_{nn}(\mathbf{k})=V_{10}+V_{11}e^{-i\mathbf{k}\cdot\mathbf{a}_1}+V_{12}e^{-i\mathbf{k}\cdot\mathbf{a}_2}+H.c.,\\
&V_{nnn}(\mathbf{k})=V_{20}+V_{21}e^{-i\mathbf{k}\cdot\mathbf{a}_1}+V_{22}e^{-i\mathbf{k}\cdot\mathbf{a}_2}+V_{23}e^{-i\mathbf{k}\cdot(\mathbf{a}_1-\mathbf{a}_2)}+H.c.,
\end{align}
where $V_{10}, \cdots, V_{23}$ are $4\times4$ matrices and are given in Appendix \ref{Vmatrix}. To obtain the energy spectrum of phonons one may use the equation of motion for field operator $\phi_{\mathbf{k}}$ \cite{Zhang:PRL2019}, $i\hbar \partial_t\phi_{\mathbf{k}}=[\phi_{\mathbf{k}}, H_{ph}]$, yielding
\begin{align}
i\hbar \partial_t\phi_{\mathbf{k}}=\eta_{ph} \tilde{H}_{ph}(\mathbf{k}) \phi_{\mathbf{k}},
\end{align}  
where $\eta_{ph}=[\phi_{\mathbf{k}}, \phi^{\dagger}_{\mathbf{k}}]=-\sigma^{y}\otimes \mathbf{1}_{4\times4}$. Therefore, the positive energy eigenvalues of $\eta_{ph} \tilde{H}_{ph}(\mathbf{k})$ will give the energy bands of phonon. Alternatively, one may unitarily transform the phonon Hamiltonian in \eqref{Hph} to bosonic creation and annihilation operators of phonons using $u^{\alpha}_{\mathbf{k}s}=(a_{\mathbf{k},\alpha s}+a^{\dagger}_{-\mathbf{k}, \alpha s})/\sqrt{2}$ and $p^{\alpha}_{\mathbf{k}s}=-i(a_{\mathbf{k},\alpha s}-a^{\dagger}_{-\mathbf{k}, \alpha s})/\sqrt{2}$, where $s=A, B$ is the sublattice index, and $a$ and $a^{\dagger}$ are, respectively, the phonon annihilation and creation operators. A paraunitary transformation is then used to Bogoliubov diagonalize the obtained bosonic Hamiltonian in particle and hole spaces.

\subsection{Magnon-phonon coupling}
We now derive the last term in \eqref{Htotal} describing the hybridization between magnon and phonon modes. To couple them, the spatial dependency of magnetic exchange interactions to the instant position of magnetic ions is taken into account. In writing down the magnetic Hamiltonian in \eqref{Hm}, we could, in principle, assume that the exchange $J_{ij}(\mathbf{R}_{ij})$ and also the DM interaction $\mathbf{D}(\mathbf{R}_{ij})$ depend on the separation between ions even when they are out of equilibrium positions. The exchanges $J_{ij}$ is expanded as,

\begin{align}
J_{ij}(\mathbf{R}_{ij})\approx J_{ij}(\mathbf{R}^{0}_{ij}) + \sum_{\alpha}\left.\frac{\partial J_{ij}}{\partial R^{\alpha}_{ij}}\right |_{\mathbf{R}^{0}_{ij}}\left(u^{\alpha}_i-u^{\alpha}_j \right).
\end{align}      

The second term depends on the vibrations of ions. However, when combined with spin exchange interaction terms $\mathbf{S}_i\cdot\mathbf{S}_j$, they generate higher order bosonic couplings such as $ab^{\dagger}b$, falling out of linear spin-wave theory used here. Therefore, we ignore such couplings and, as discussed in Sec.\ref{Sec:magnetic}, the magnetic exchange interactions $J$ and $J'$ are assumed to have the same values as in equilibrium.      

The DM interaction could in general have both out-of-plane and in-plane components, $\mathbf{D}(\mathbf{R}_{ij})=D_z(\mathbf{R}_{ij})\hat{z}+\mathbf{D}_{\parallel}(\mathbf{R}_{ij})$, with $\hat{z}\cdot \mathbf{D}_{\parallel}(\mathbf{R}_{ij})=0$. The direction of DM vector is restricted by lattice  symmetry\cite{DM1,DM2} . For exchange path between second neighbors on the honeycomb lattice, the out-of-plane component, $D_z(\mathbf{R}_{ij})\hat{z}$ is allowed by symmetry as this path is locally asymmetric. On the honeycomb lattice as shown in Fig.~\ref{lattice}, for the path between A sites, the site B is located on the left or right of the paths along the $\mathbf{a}_1$ and $\mathbf{a}_2$, respectively. Therefore, this exchange component naturally occurs in honeycomb lattice. However, and in parallel arguments made above for $J_{ij}$, the expansion of $D_z(\mathbf{R}_{ij})$ in displacements will generate higher-order interactions between magnons and phonons. Thus, we only consider the equilibrium distribution of $D_z(\mathbf{R}^{0}_{ij})$. Also note that for a pristine honeycomb lattice the first neighbor bonds are symmetric and therefore the $D_z(\mathbf{R}_{ij})\hat{z}$ is identically zero. However, upon breaking the mirror symmetry w.r.t the plane, say by growing the 2D honeycomb lattice on a substrate, an in-plane component arises. That is $\mathbf{D}_{\parallel}(\mathbf{R}_{ij})=-D_{\parallel}(\mathbf{R}_{ij}) \hat{z}\times\hat{R}_{ij}$. In Fig.~\ref{lattice} the directions of $\mathbf{D}_{\parallel}$ on first neighbor bonds are shown. Note that the latter vector will not affect the magnon spectrum within the linear spin wave theory, as considered in Eq.\eqref{Hm}.
For example, let us consider $D_x(\mathbf{S}_i\times\mathbf{S}_j)_x$. Up to first order in magnon operators, $D_x(S^{x}_iS^{z}_j-S^{z}_iS^{x}_j)\propto (b_i+b_i^{\dagger}-b_j-b_j^{\dagger})$, and thus it vanishes by sum over all sites. 

We shall however argue that the spatial dependency of $\mathbf{D}_{\parallel}(\mathbf{R}_{ij})$ will generate magnon-phonon coupling\cite{Zhang:PRL2019}. Doing so, let us write the spin operators as $\mathbf{S}_i=S\hat{z}+\delta\mathbf{S}_i$, where the first term is the magnetic ordering of the classical ground state and $\delta\mathbf{S}_i$ describes the fluctuations around it. The in-plane DM interaction is then cast into,
\begin{align}
\mathbf{D}_{\parallel}(\mathbf{R}_{ij})\cdot(\mathbf{S}_i\times\mathbf{S}_j)=D_{\parallel}(\mathbf{R}_{ij})S(\delta\mathbf{S}_i-\delta\mathbf{S}_j)\cdot\hat{R}_{ij}. 
\end{align} 

Coupling to phonons arises by expanding $\mathbf{D}_{\parallel}(\mathbf{R}_{ij})$ in its argument, $\mathbf{R}_{ij}=\mathbf{R}^{0}_{ij}+\mathbf{u}_{i}-\mathbf{u}_{j}$, around equilibrium positions up to first order in displacements. We also expand the unit vector $\hat{R}_{ij}$ to the same order. The magnon-phonon coupling Hamiltonian reads as\cite{Zhang:PRL2019}  

\begin{align}\label{Hmph}
H_c=\sum_{<ij>}\left(u^{\alpha}_i-u^{\beta}_j\right)T^{\alpha\beta}_{ij}\left(\delta S^{\alpha}_i-\delta S^{\beta}_j\right),
\end{align}        
where
\begin{align}
T^{\alpha\beta}_{ij}=\frac{D_{\parallel}S}{R}\left[\delta^{\alpha\beta}-(1+\gamma)\hat{R}^{0\alpha}_{ij}\hat{R}^{0\beta}_{ij} \right]
\end{align}
with $D_{\parallel}=D_{\parallel}(\mathbf{R}_{ij}^{0})$ and $\gamma=-(dD_{\parallel}/dR)(R/D_{\parallel})$. In momentum space the coupling Hamiltonian \eqref{Hmph} takes the following form,

\begin{align}
H_c=\sum_{\mathbf{k}}\phi^{\dagger}_{ph}(\mathbf{k}) H_{c}(\mathbf{k})\phi_{m}(\mathbf{k}),
\end{align}
where $\phi_{ph}(\mathbf{k})=(u^x_{\mathbf{k}A}, u^y_{\mathbf{k}A}, u^x_{\mathbf{k}B}, u^y_{\mathbf{k}B})^{t}$ and $\phi_{m}(\mathbf{k})=(\delta S^x_{\mathbf{k}A}, \delta S^y_{-\mathbf{k}A}, \delta S^x_{\mathbf{k}B}, \delta S^y_{-\mathbf{k}B})^{t}$ group the displacements' and magnetic fluctuations' fields. Note that in terms of magnon operators $\delta S^{x}_{\mathbf{k}}=\sqrt{S/2}(b_{\mathbf{k}} + b^{\dagger}_{-\mathbf{k}})$ and  $\delta S^y_{-\mathbf{k}}=-i\sqrt{S/2}(b_{\mathbf{k}} - b^{\dagger}_{-\mathbf{k}})$. The coupling $H_{c}(\mathbf{k})$ is written as,
\begin{align}\label{Hcoupling}
H_{c}(\mathbf{k})=
\begin{pmatrix}
  T_{\mathbf{k}=0} & -T_{\mathbf{k}} \\
  -T_{-\mathbf{k}} & T_{\mathbf{k}=0}  
 \end{pmatrix}.
\end{align} 

Here, the entities are matrices in sublattice basis: $T_{\mathbf{k}}=T_0+T_1e^{-i\mathbf{k}\cdot\mathbf{a}_1}+T_2e^{-i\mathbf{k}\cdot\mathbf{a}_2}$, where 

\begin{align}\nonumber
&T_0=\zeta SD_{\parallel}\begin{pmatrix}
  1 & 0 \\
  0 & -\gamma  
 \end{pmatrix},\\ \nonumber
&T_1=\frac{\zeta SD_{\parallel}}{4}\begin{pmatrix}
  1-3\gamma & -\sqrt{3}(1+\gamma) \\
  -\sqrt{3}(1+\gamma) & 3-\gamma  
 \end{pmatrix},\\
&T_2=\frac{\zeta SD_{\parallel}}{4}\begin{pmatrix}
  1-3\gamma & \sqrt{3}(1+\gamma) \\
  \sqrt{3}(1+\gamma) & 3-\gamma  
 \end{pmatrix},
\end{align}
where $\zeta=(1/R)\sqrt{\hbar/2M\Omega}$ is a dimensionless quantity and we set $\gamma=0$ for simplicity. By a proper choices of parameters relevant to materials, $\zeta\simeq0.1$. Now we can write the total Hamiltonian in \eqref{Htotal} as 

\begin{align}\label{HtotalK}
H=\frac{1}{2}\sum_{\mathbf{k}}\tilde{\psi}^{\dagger}_{\mathbf{k}}\tilde{H}(\mathbf{k})\tilde{\psi}_{\mathbf{k}}
\end{align}
where $\tilde{\psi}_{\mathbf{k}}=(\delta S^x_{\mathbf{k}A}, \delta S^y_{-\mathbf{k}A}, \delta S^x_{\mathbf{k}B}, \delta S^y_{-\mathbf{k}B}, u^{x}_{\mathbf{k}A}, u^{y}_{\mathbf{k}A}, u^{x}_{\mathbf{k}B}, u^{y}_{\mathbf{k}B},\\ p^{x}_{-\mathbf{k}A}, p^{y}_{-\mathbf{k}A}, p^{x}_{-\mathbf{k}B}, p^{y}_{-\mathbf{k}B} )^t$ and 
\begin{align}\label{HfullK}
\tilde{H}(\mathbf{k})=\begin{pmatrix}
  \tilde{H}_m(\mathbf{k}) & \tilde{H}_c(\mathbf{k})\\
  \tilde{H}^{\dagger}_c(\mathbf{k}) & \tilde{H}_{ph}(\mathbf{k}) 
 \end{pmatrix}.
\end{align}

Here, $\tilde{H}_{m}(\mathbf{k})$ is obtained from \eqref{HmK} by adding a hole space to the particle space, i.e.,
\begin{align}
\tilde{H}_m(\mathbf{k})=\begin{pmatrix}
  H_m(\mathbf{k}) & \mathbf{0}_{2\times2}\\
  \mathbf{0}_{2\times2} & H^{T}_m(-\mathbf{k})
 \end{pmatrix},
\end{align}
and 
\begin{align}
\tilde{H}_c(\mathbf{k})=\begin{pmatrix}
  H^{\dagger}_c(\mathbf{k}) & \mathbf{0}_{4\times4}
 \end{pmatrix}.
\end{align} 
            
The Bloch Hamiltonian $\tilde{H}(\mathbf{k})$ in \eqref{HtotalK} yields a full account of band structure of the hybrid magnon-phonon modes, and the effects of magnetoelasticity on the energy bands and the transport properties can be readily studied. The energy bands are obtained by diagonalizing $\tilde{\eta} \tilde{H}(\mathbf{k})$ with $\tilde{\eta}=[\psi_{\mathbf{k}}, \psi^{\dagger}_{\mathbf{k}}]$, the positive eigenvalues yield the energies and the negative ones are redundant.      

\section{Magnetoelastic spectrum and Berry curvature: the honeycomb lattice\label{BerryHoney}}

\subsection{Energy bands}
\begin{figure}[t]
   \center
    \includegraphics[width=\linewidth]{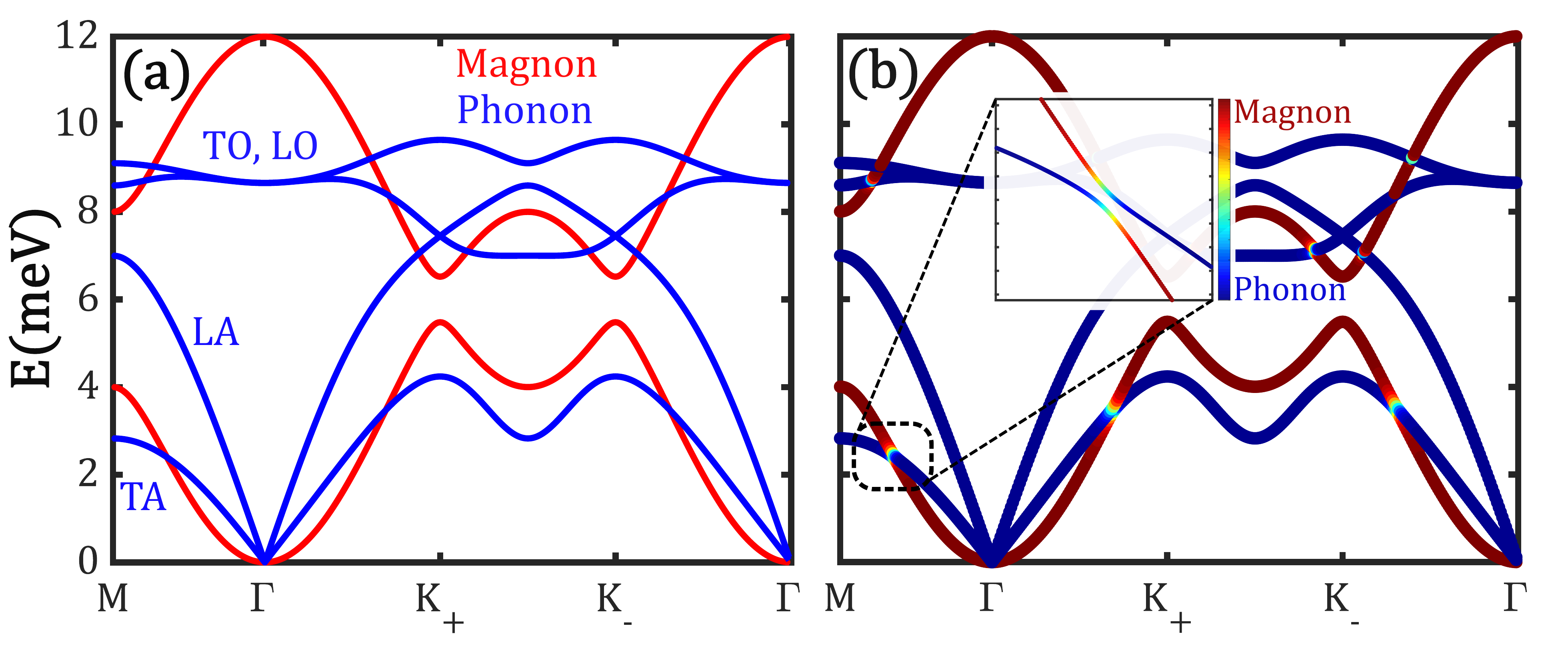}
    \caption{ (a) Energy bands of magnons (solid red lines) and phonons (solid blue lines) in the absence of in-plane DM interaction. The magnons energy bands are gaped due to out-of-plane DM interaction $D_z$ endowing them with nontrivial band topology. The in-plane vibrational modes of two ions lead to four phonon branches: transverse acoustic (TA), longitudinal acoustic (LA), transverse optical (TO), and longitudinal optical (LO). (b) The hybridized magnon-phonon energy bands in the presence of in-plane DM interaction $D_{\parallel}$. The inset indicates an avoided crossing between the magnon (red) and phonons (blue) along $\Gamma-M$ direction. The parameters are $J=2$ meV, $D_z=0.1$ meV, $D_{\parallel}=0.5$ meV, $\hbar\Omega=5$ meV, $\hbar\Omega'=2$ meV. } \label{honey:bands}
\end{figure}

The band structure of the honeycomb model \eqref{HtotalK} along the high-symmetry lines of the Brillouin zone is shown in Fig.~\ref{honey:bands}. In the absence of magnon-phonon coupling the phonon dispersion is depicted by solid blue lines and the magnon dispersion by solid red lines as shown in Fig.~\ref{honey:bands} (a). As explained in Sec.~\ref{Sec:magnetic}, the latter bands are topologically nontrivial characterized by nonzero Chern numbers $c=\pm1$. The phonon spectrum consists of four energy bands, two acoustic and two optical modes being consistent with having two sublattices and $x, y$ displacements of each magnetic ion. In the absence of second neighbor ionic interactions, $\Omega'=0$ in \eqref{HphR}, the transverse acoustic (TA) mode becomes completely flat with zero energy over the entire BZ. Also, the transverse optical (TO) branch becomes flat but at finite energy corresponding to the frequency $\Omega$. The former TA branch form a set of deformation modes of zero energy, the so-called floppy modes, due to small coordination number below the isostatic point\cite{pnas:sun2012}. The number of zero modes are consistent with Maxwell criterion. According to this criterion, for a $d$-dimensional lattice with $N$ sites and coordination $z<2d$, the number of zero modes is $N_0=dN-\frac{1}{2}zN$ giving rise to $N_0=\frac{1}{2}N$ for the honeycomb lattice. This is equal to the number of unit cells, i.e., the number of $k$ points in BZ. Of this huge set of zero modes, two modes are trivial associated with the rigid translations and the remaining modes are internal floppy modes\cite{pnas:sun2012}. Therefore, by adding the second neighbor interactions, i.e., by increasing the coordination number, the zero deformation modes become dispersive and acquire finite energy. Hence, the lattice becomes an elastic solid and the mechanical stability is achieved. We shall derive an effective description of low-energy bands where the dependency on $\Omega'$ becomes manifest. 

In Fig.~\ref{honey:bands} (b) we show the energy bands in the presence of in-plane DM interaction $D_{\parallel}$. For the parameters chosen, the magnon bands are hybridized with the phonons significantly. The lower magnon band only crosses the TA branch\cite{Takahashi:PRL2016}. At crossings, the DM interaction hybridizes the bands causes energy splitting. For the lowest bands along the $\Gamma-M$ direction,  the avoided crossing is shown in the inset. It is clearly seen that the wave function contents of the bands change from purely magnons to phonons and vice versa. We will show that such avoided crossings generate a new Berry curvature in addition to the intrinsic Berry curvature of magnons bands.          

In order to understand the band hybridization in the vicinity of avoided crossings, we develop an effective model of magnon-phonon hybridization. For sake of simplicity we consider the crossing along the $\Gamma-M$ direction as shown in the inset of Fig.~\ref{honey:bands} (b). In the coordinate system describing the ion vibrations in the $x-y$ plane ($u_x-u_y$ vibrations), of all symmetry related $\Gamma-M$ directions we choose the one along the $y$ axis. Therefore, the transverse modes will have vibrations along the $x$-axis. In this restricted subspace the inter-ionic interaction reads
\begin{align}\nonumber
V_{T}=&\frac{1}{2}\sum_{k}\Big\{\left[3M\Omega^2+2M\Omega'^2(1-\cos k) \right]\left(u^{x}_{-kA}u^{x}_{kA}+u^{x}_{-kB}u^{x}_{kB} \right)\\  \label{Veff}
&-3M\Omega^2\left(e^{-ik}u^{x}_{-kA}u^{x}_{kB}+ e^{ik}u^{x}_{-kB}u^{x}_{kA}\right)\Big\}.
\end{align}

Using the following transformation to transverse modes,
\begin{align}
u_{TA}(k)=\frac{1}{\sqrt{2}}\left(e^{ik}u^{x}_{kA}+u^{x}_{kB} \right),~u_{TO}(k)=\frac{1}{\sqrt{2}}\left(-e^{ik}u^{x}_{kA}+u^{x}_{kB} \right),
\end{align} 
the phonon Hamiltonian including the interaction \eqref{Veff}  becomes, 
\begin{align}\nonumber 
H_{ph, T}=&\frac{1}{2M}\sum_{k}\left[|p_{TA}(k)|^2+ |p_{TO}(k)|^2\right]\\ \label{Hph_eff}
&+\frac{M}{2}\sum_{k}\left[ \Omega^2_{TA}(k)|u_{TA}(k)|^2 +\Omega^2_{TO}(k)|u_{TO}(k)|^2\right],
\end{align}
where we define the acoustic and optical dispersion frequencies as 
\begin{align}
&\Omega_{TA}(k)=\Omega'\left[1-\cos k\right]^{1/2},\\
&\Omega_{TO}(k)=\Omega\left[3+\xi'^2(1-\cos k)\right]^{1/2}. 
\end{align}
 
\begin{figure}[t]
   \center
    \includegraphics[width=\linewidth]{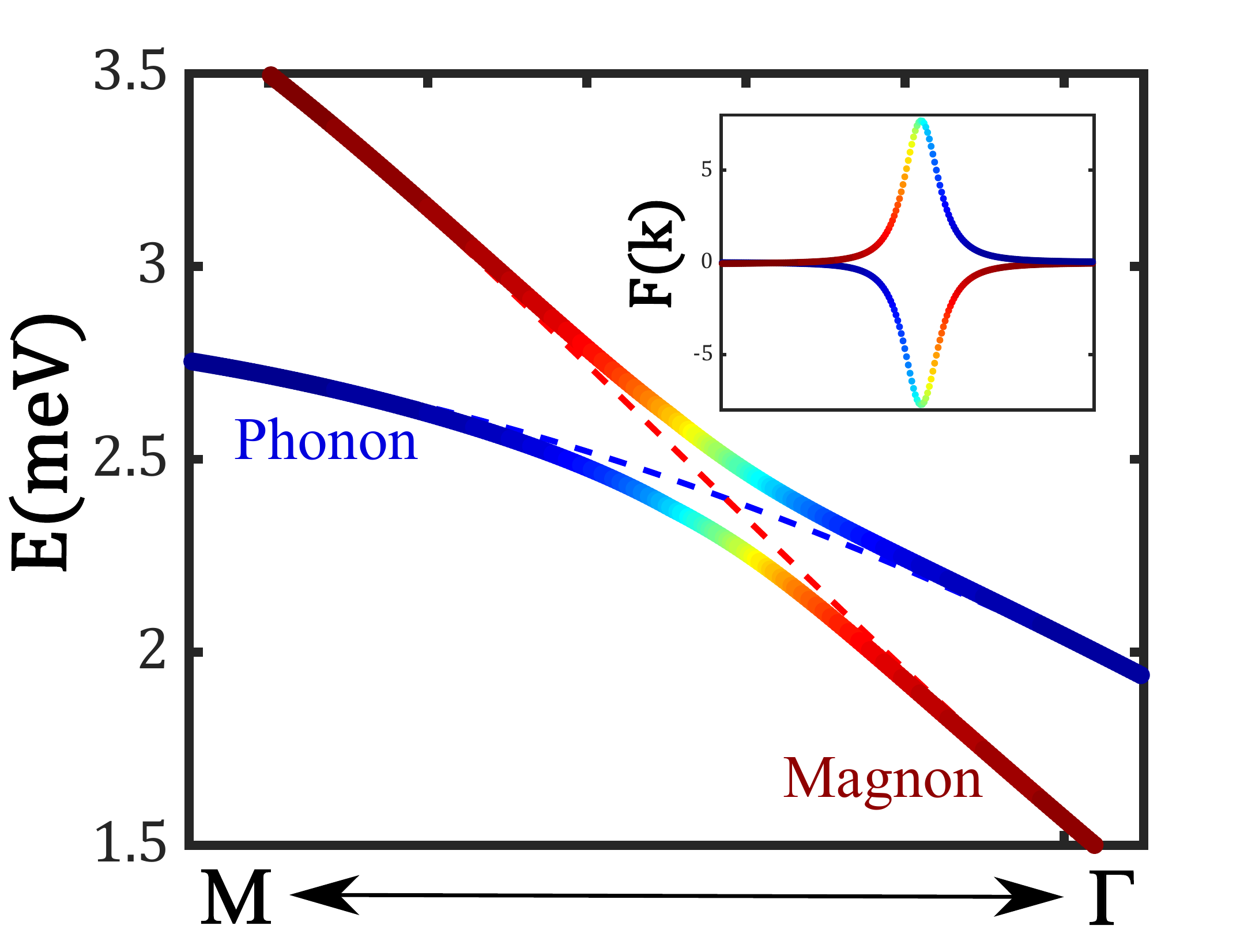}
    \caption{Main panel: energy bands of effective magnon-phonon model in \eqref{hk}, around the avoided crossing, along the $\Gamma-M$ direction. The blue and red dashed lines are energy dispersion of phonons $\varepsilon_{ph,k}$ and magnons $\varepsilon_{m,k}$, respectively, in the absence of hybridization $\Delta_k$ (i.e., $D_{\parallel}=0$). The crossing is avoided with in-plane DM interaction shown by thick lines. Inset: Berry curvatures of hybridized bands along the same direction. The color spectrum and the parameters are the same as in Fig.~\ref{honey:bands}.} \label{honey:effective}
\end{figure}

The transverse phonons become dispersionless for $\Omega'=0$ ($\xi'=\Omega'/\Omega=0$). In particular, the dispersion vanishes for the TA branch, implying that a second neighbor inter-ion potential energy is required to have a stable lattice. Since we are interested in low-energy part of the spectrum, below we only consider the TA modes in \eqref{Hph_eff}. Changing the displacement and momentum to dimensionless variables, $u_{TA}(k)=\sqrt{\hbar/M\Omega_{TA}(k)}u_k$ and $p_{TA}(k)=\sqrt{\hbar M \Omega_{TA}(k)}p_k$ and using the transformation $u_k=\left(a_k+a^{\dagger}_{-k} \right)/\sqrt{2}$ and $p_k=\left(a_k-a^{\dagger}_{-k} \right)/\sqrt{2}i$, the transverse phonon Hamiltonian can be written as,
\begin{align}\label{Hph_eff2}
H_{ph, T}=\sum_{k}\varepsilon_{ph, k}\left(a^{\dagger}_{k}a_k +\frac{1}{2}\right)
\end{align}  
where $\varepsilon_{ph,k}=\hbar\Omega_{TA}(k)$. 

The lower magnon band is described by the following Hamiltonian,
\begin{align}\label{HmL}
H_{m, l}=\sum_{k}\varepsilon_{m,k}b^{\dagger}_{k,-}b_{k,-},
\end{align}
where $\varepsilon_m(k)=3SJ-JS|f(k)|$ is the magnon dispersion with $f(k)=1+2e^{-ik}=|f(k)|e^{-i\theta_k}$, and magnon creation (annihilation) operators $b^{\dagger}_{k,-}$ ($b_{k,-}$) describes the projection to the lowest magnon band indicated by subindex ``-". In the following, we drop the subindex and identify $b_{k}\equiv b_{k,-}$. Finally, by projecting the magnon-phonon coupling Hamiltonian \eqref{Hcoupling} to lowest bands, we obtain
\begin{align}\label{HcL}
H_{c, l}=\sum_{k}g_{k}u_{TA}(-k)\left(b_{k}+b^{\dagger}_{-k}\right),
\end{align}
where 
\begin{align}
g_{k}=\frac{1}{2\sqrt{2}}\frac{SD_{\parallel}}{R}\left(1-e^{ik}\right)\left(1-e^{-i\theta_k} \right).
\end{align}

Using \eqref{Hph_eff2}, \eqref{HmL}, and \eqref{HcL}, the low-energy modes are described by the following effective Hamiltonian,
\begin{align}
H_{eff}=\frac{1}{2}\sum_{k}\varphi^{\dagger}_{k}H_{eff}(k)\varphi_{k}.
\end{align} 

Here, the boson creation and annihilation operators are grouped in $\varphi_{k}=(a_k, b_k, a^{\dagger}_{-k}, b^{\dagger}_{-k})^t$, and              
\begin{align}\label{Heff}
H_{eff}(k)=\begin{pmatrix}
  h_k & M_k\\
 M^{\dagger}_k & h_{-k}^{t} 
 \end{pmatrix},
\end{align}
where
\begin{align}\label{hk}
h_k=\begin{pmatrix}
  \varepsilon_{ph,k} & \Delta_k\\
  \Delta_{-k} & \varepsilon_{m,k}
 \end{pmatrix},~~
M_k=\begin{pmatrix}
  0 & \Delta_k\\
  \Delta_{-k} & 0
 \end{pmatrix}.
\end{align}

The magnon-phonon band hybridization is given by $\Delta_k$:
\begin{align}\label{Delta}
\Delta_k=\frac{\zeta g_{k}}{(1-\cos k)^{1/4}},
\end{align} 
where $\zeta=(1/R)\sqrt{\hbar/2M\Omega'}$ is a dimensionless parameter. In Fig.~\ref{honey:effective} we plot the energy spectrum of effective Hamiltonian \eqref{Heff}. Besides the hybridized energy bands, we also depict the bare energy bands in the absence of hybridization $\Delta_k$ shown by dashed lines. The spectrum demonstrates that the effective model \eqref{Heff} readily reproduces the spectrum shown in the inset of Fig.~\ref{honey:bands}, where we used the full Hamiltonian \eqref{HfullK}. We also found that if the particle-hole coupling matrix $M_k$ is manually set to zero, i.e., we neglect the product of two creation and annihilation operators, the spectrum of $H_{eff}(k)$ is continuously connected to spectrum generated by two-band model $h_k$ in eq.~\eqref{hk} without gap closing; the $M_k$ just changes the values of energies slightly. Therefore, we assert that the hybridization of magnon and phonons can be described by $h_k$ and the splitting is triggered off by $\Delta_k$ in \eqref{Delta}. The color density indicates the weights of magnons and phonons in the wave functions. Moving along one of the bands the hybridization $\Delta_k$ winds the wave function form magnons to phonons and vice versa.

\subsection{Berry curvature of hybridized bands}
As discussed in Sec.\ref{Sec:magnetic}, the pure magnon bands are topologically nontrivial characterized by Chern number $c=-1$ for the lower band and $c=+1$ for the higher band, respectively. The pure phonon bands, on the other hand, are topologically trivial. We show that the avoided crossings caused by the in-plane DM interaction endows the hybridized bands with nonzero Berry curvature. 

To compute the Berry curvature of the model \eqref{HfullK}, we use the following expression\cite{Matsumoto:PRB2014}

\begin{align}\label{Berry1_formula}
F_n(\mathbf{k})=i\epsilon_{\mu\nu}\left[\eta\partial_{\mu}T^{\dagger}_{\mathbf{k}} \eta\partial_{\nu}T_{\mathbf{k}} \right]_{nn},
\end{align}     
where $T_{\mathbf{k}}$ is a paraunitary transformation used to obtain the spectrum of a bosonic Hamiltonian in the particle-hole space and $\partial_{\mu}T_{\mathbf{k}}=\partial T_{\mathbf{k}}/\partial k_{\mu}$. For numerical calculations we rewrite the above expression into a more convenient form. We unitarily transform the $\tilde{H}(\mathbf{k})$ in \eqref{HfullK} to $H(\mathbf{k})=U^{\dagger}\tilde{H}(\mathbf{k})U$, where $U$ is the corresponding unitary transformation $\tilde{\psi}_{\mathbf{k}}=U\psi_{\mathbf{k}}$ to a new basis: $\psi_{\mathbf{k}}=(b_{\mathbf{k}A}, b_{\mathbf{k}B}, a^{x}_{\mathbf{k}A}, a^{y}_{\mathbf{k}A}, a^{x}_{\mathbf{k}B}, a^{y}_{\mathbf{k}B}, b^{\dagger}_{-\mathbf{k}A}, b^{\dagger}_{-\mathbf{k}B}, a^{x \dagger}_{-\mathbf{k}A}, a^{y\dagger}_{-\mathbf{k}A}, a^{x\dagger}_{-\mathbf{k}B}, a^{y\dagger}_{-\mathbf{k}B})^t$. Doing so, the paraunitary transformation satisfies\cite{Shindou:PRB2013} 
\begin{align}
T_{\mathbf{k}}\eta T^{\dagger}_{\mathbf{k}}=\eta,~~~H(\mathbf{k})T_{\mathbf{k}}=\eta T_{\mathbf{k}} \varepsilon_{\mathbf{k}},
\end{align}    
where $\eta$ is a diagonal matrix with $+1$ for particle space and $-1$ for hole space, and $\varepsilon_{\mathbf{k}}$ is a matrix of energy eigenvalues with values  $E_{\mathbf{k}}>0$ for particles and $-E_{-\mathbf{k}}<0$ for holes. Taking the momentum derivative of the above eigenvalue problem, we obtain the following matrix element,
\begin{align}
\langle n|\partial_{\mu}T_{\mathbf{k}}|m\rangle=-\frac{\langle n|\eta \bar{V}_{\mu\mathbf{k}}|m\rangle}{(\varepsilon_{\mathbf{k}})_{nn}-(\varepsilon_{\mathbf{k}})_{mm}},
\end{align} 
where $\bar{V}_{\mu\mathbf{k}}=T^{\dagger}_{\mathbf{k}}\partial_{\mu}H(\mathbf{k})T_{\mathbf{k}}$. Using this matrix element and $\eta T^{\dagger}_{\mathbf{k}}=T^{-1}_{\mathbf{k}}\eta$, the Berry curvature \eqref{Berry1_formula} is cast as\cite{Shindou2:PRB2013, Hwang:PRL2020} 

\begin{align}\label{Berry2_formula}
F_{n}(\mathbf{k})=i\epsilon_{\mu\nu} \sum_{m\neq n}\frac{\langle n|\eta \bar{V}_{\mu\mathbf{k}}|m\rangle \langle m|\eta \bar{V}_{\nu\mathbf{k}}|n\rangle}{\left[(\varepsilon_{\mathbf{k}})_{nn}-(\varepsilon_{\mathbf{k}})_{mm} \right]^2}.
\end{align}


\begin{figure}[t]
   \center
    \includegraphics[width=\linewidth]{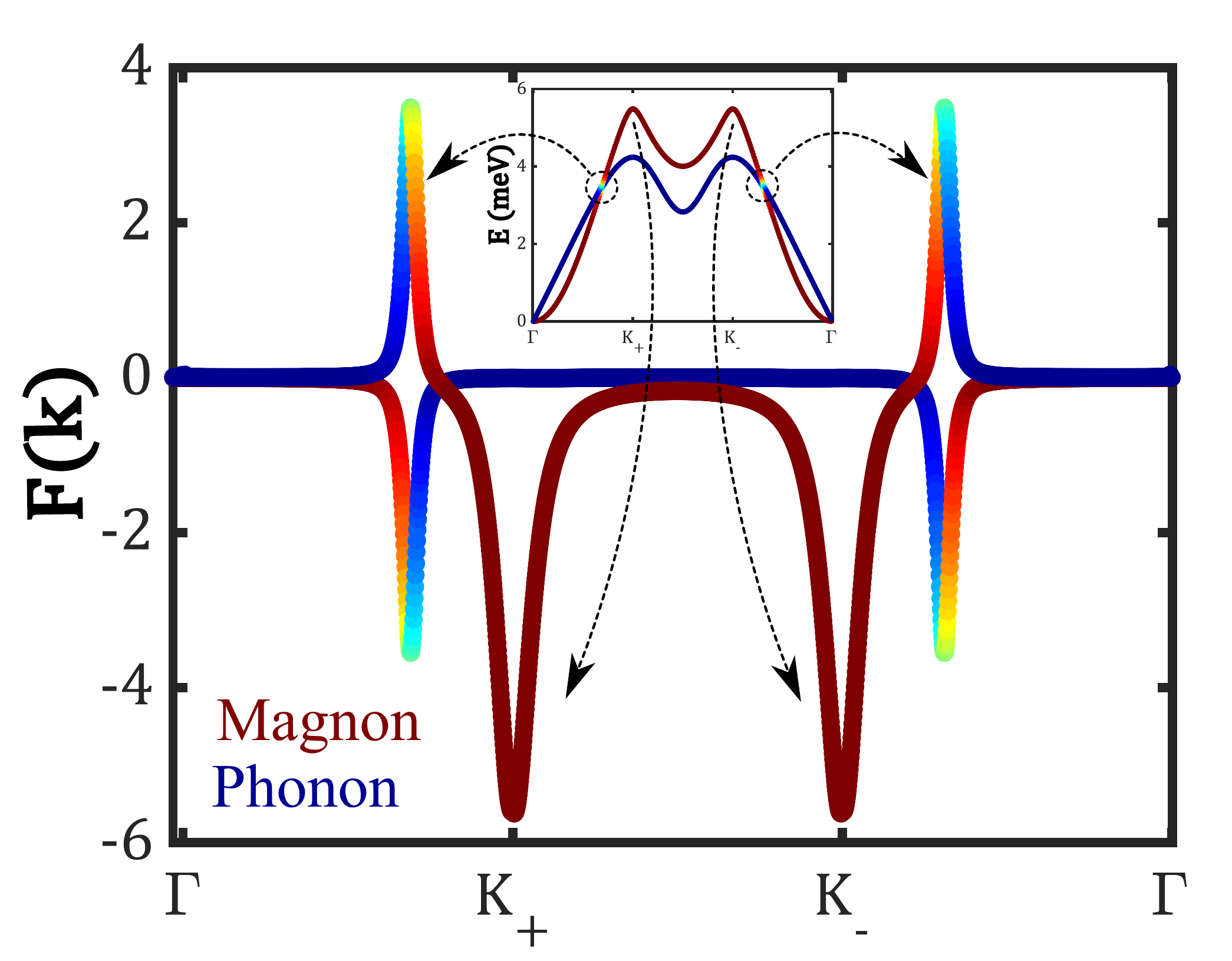}
    \caption{Main panel: Berry curvature of two lowest magnetoelastic energy bands along the high-symmetry lines $\Gamma-\mbox{K}_{+}-\mbox{K}_{-}-\Gamma$. Inset: the corresponding lowest energy bands. Two negative pronounced peaks in the main panel are associated with massive Dirac cones of magnon bands shown by long curved dashed arrows. Besides, new peaks appear in Berry curvature resulting from the hybridized magnon-phonon bands. Color spectrum indicates the magnon (the reddest one) versus phonon (the bluest one) contributions of the Berry curvature. The parameters used are the same as in Fig.~\ref{honey:bands}} \label{Berry1}
\end{figure}

In the inset of Fig.~\ref{honey:effective} we plot Berry curvature \eqref{Berry2_formula} along the $\Gamma-M$ direction. When the hybridization is set to zero, for which the energy bands are shown by dashed lines,  the Berry curvature of magnons along this direction is nearly zero. Note that main contribution to the Berry curvature of magnons result from the region neat the Dirac points $\mathrm{\mbox{K}}_{\pm}$ resulting to nontrivial band topology \cite{Owerre_2016,Kim:PRL2016}. The inset clearly indicates that upon hybridization of magnon and phonons, the Berry curvature of states near the avoided crossing is strongly increased: more the states are hybridized, as shown by light colors, more the corresponding Berry curvatures are pronounced.        

\begin{figure}[t]
   \center
    \includegraphics[width=\linewidth]{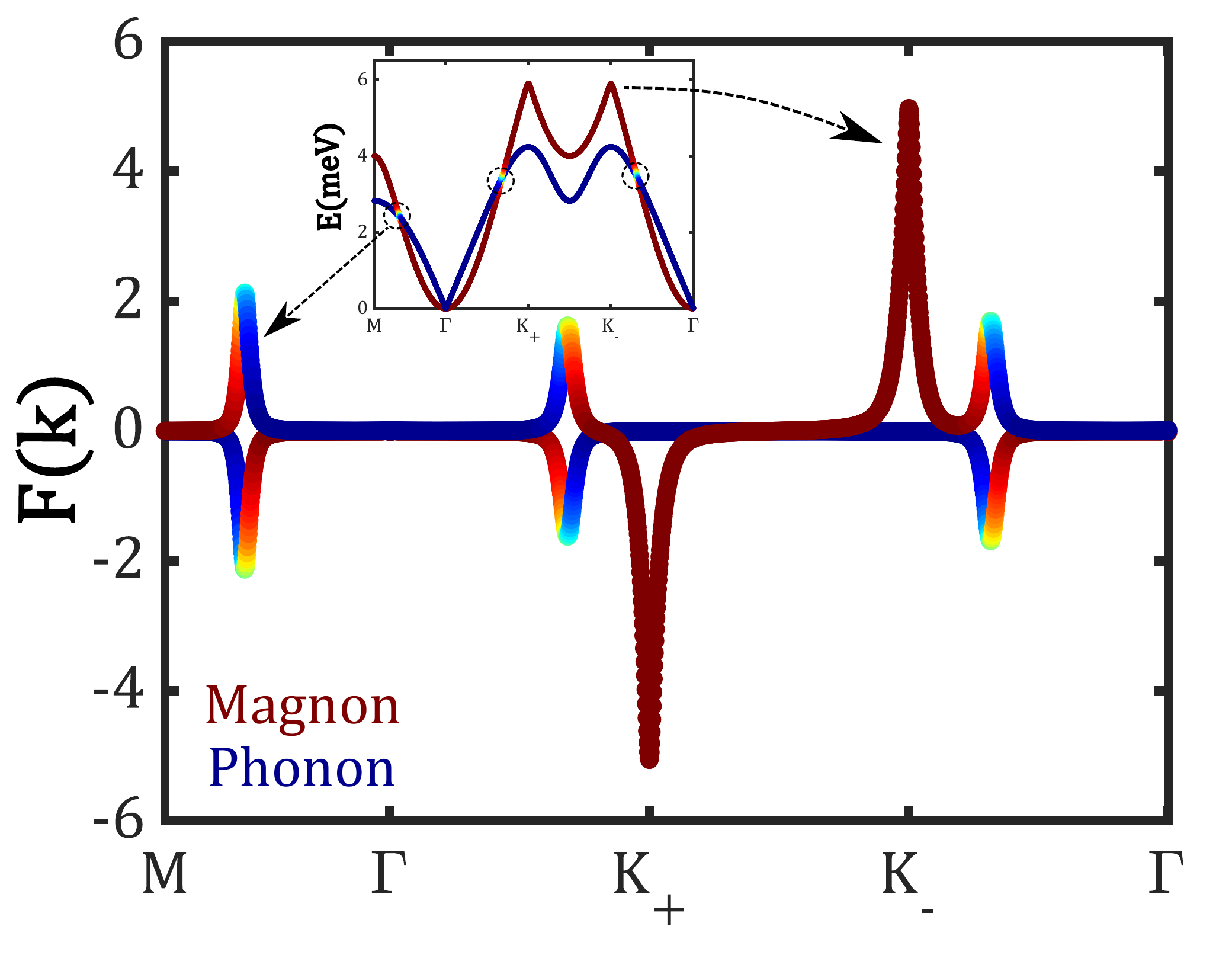}
    \caption{The same as Fig.~\ref{Berry1} but with trivial magnon bands. Inset shows the two lowest energy bands. The magnon spectrum around $\mbox{K}_{-}$ now has positive Berry curvature while the spectrum around $\mbox{K}_{+}$ has negative values. The crossings in magnetoelastic bands shown by dashed circles lead to peaks in Berry curvatures around anticrossing points. Note that here the numbers on vertical axis are $\mathrm{sgn}(F(\mathbf{k}))\ln(1+|F(\mathbf{k})|)$.} \label{Berry2}
\end{figure}

Fig.~\ref{Berry1} shows the Berry curvature of two lowest energy bands (shown in the inset) along a direction including Dirac points and avoided crossings. Two pronounced negative peaks around $\mathrm{\mbox{K}}_{\pm}$ in the Berry curvature are associated with the band topology of magnons. Note that these peaks exist even in the absence of phonons, and as pointed out before, when integrated over the entire BZ, it yields Chern number $c=-1$ for the lower magnon band. The band hybridization adds new features to the Berry curvature near the crossings. The locations of avoided crossings are shown by dashed circles in the inset and the small curved arrows point to the corresponding peaks in the Berry curvature of lower and upper bands. As we pointed out above, in the absence of hybridization the Berry curvature is mainly distributed around the Dirac nodes of magnon located at energies higher than the phonons. Hybridization redistributes the Berry curvature by creating new peaks near the avoided crossings. Now the lower band is characterized by the sole negative Berry curvature, whilst the Berry curvature of the upper band acquires both positive (due to hybridization to phonons) and negative (due to sole magnons) contributions. Therefore, when integrated over the entire BZ, the Chern number of lower band is $-1$ and for the upper band is $0$. This redistribution of Berry curvature from the upper to lower bands will have important implications on the thermal Hall conductivity.

Next we consider an interesting case of trivial magnon energy bands. The latters are achieved by considering a ferromagnetic state with slightly different magnetizations on sublattices, e.g., the ferrimagnet Fe$_2$Mo$_3$O$_8$\cite{Park:Nano2020} with different magnetic ions on sublattices. The Holstein-Primakoff transformation yields $S_{i}^{z}S_{j}^{z}\simeq -S_i b^{\dagger}_{j}b_{j}-S_j b^{\dagger}_{i}b_{i}$. Taking the classical magnetization to be $S_i=S+\delta S$ and $S_j=S-\delta S$, a sublattice potential term, $J\delta S(b^{\dagger}_{i}b_{i}-b^{\dagger}_{j}b_{j})$, is added to the magnetic Hamiltonian \eqref{Hmag}. This is equivalent to the sublattice potential for spinless electrons in the Haldane model. In the absence of normal DM interaction, i. e., $D_z=0$, the magnons bands are trivial with zero Chern numbers. Yet, the magnetoelastic energy bands arise in the presence of in-plane DM interaction $D_{\parallel}$ and, interestingly, leads to topological bands. Fig.~\ref{Berry2} shows the profile of Berry curvature of two lowest magnetoelastic bands. In contrast to the topological magnons bands in Fig.~\ref{Berry1}, the magnon wave functions near the Dirac points have opposite Berry curvatures resulting in zero Chern number. However, when hybridized with phonons, the lower band acquires negative Berry curvatures near the avoided crossings and the upper band would develop positive values. That is, the lower magnetoelastic band is characterized by Chern number $-1$ and the upper one by $+1$. This means that the topological bands emerge out of trivial bands by proper hybridization due to $D_{\parallel}$. A similar observation has also been reported for the spin-plasma modes in magnetic topological interfaces\cite{Efimkin:arXiv2020}. Since the magnetoelastic bands, as shown in the inset, have different energy dispersions, the collective modes of the hybrid system yield a finite thermal Hall conductivity response \cite{Zhang:PRL2019,Park:Nano2020, Efimkin:arXiv2020}. We discuss it in the following subsection.

\subsection{Thermal Hall Conductivity: the Honeycomb lattice}
The magnons and phonons are electrically neutral, calling for thermal Hall responses to diagnose their nontrivial Berry curvatures and band topology. For bosons the thermal Hall conductivity $\kappa_{xy}$, which measures the transverse heat current density in response to an applied temperature gradient as $J^{Q}_{xy}=\kappa_{xy}(-\boldsymbol{\nabla}T)_{y}$, is described as\cite{Matsumoto:PRL2011, Matsumoto:PRB2011, Matsumoto:PRB2014} 

\begin{align}\label{kappa}
\kappa _{xy}=- \frac{k_B^2T}{\hbar \mathcal{A}}\sum_{\mathbf{k}}\sum_{n=1}^{n=N}\left\{{c_2}[g(E _{n\mathbf{k}})]-\frac{\pi^2}{3} \right\}F_{n}(\mathbf{k}), 
\end{align}
where $\mathcal{A}$ is the area of the system, $k_B$ is the Boltzmann constant,  $g(\varepsilon)=\left(e^{\varepsilon/k_{B}T}-1\right)^{-1}$ is the Bose-Einstein distribution function, $E_{n\mathbf{k}}$ is the energy of the magnetoelastic waves, and $c_{2}(x)=(1 + x)\left(\ln \frac{1 + x}{x}\right)^2 - (\ln x)^2 - 2\mathrm{Li}_{2}(- x)$ with $\mathrm{Li}_{2}(x)$ as the polylogarithm function of second order. 

\begin{figure}[t]
   \center
    \includegraphics[width=\linewidth]{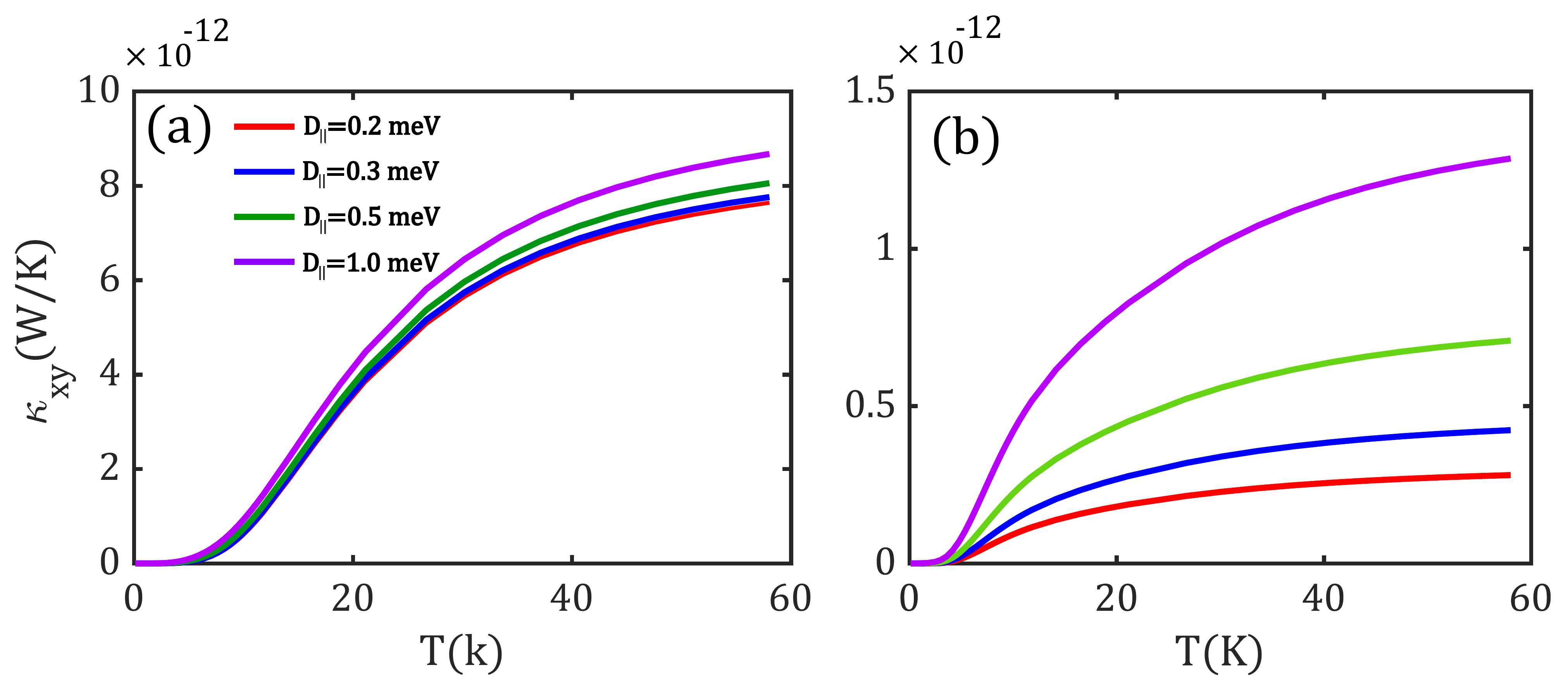}
    \caption{ The temperature dependence of thermal Hall conductivity $\kappa_{xy}$ for different values of in-plane DM interaction $D_{\parallel}$. In (a) we set $D_z=0.1$ meV: the magnon spectrum is topological. In (b) the magnon spectrum is made topologically trivial by setting $D_z=0$ and sublattice potential $J\delta S=0.1$ meV. The magnetoelastic bands arising from $D_{\parallel}$, otherwise trivial bands, carry finite thermal Hall response. } \label{Hall_honey}
\end{figure}

Fig.~\ref{Hall_honey} shows the temperature dependency of the thermal Hall conductivity for various values of in-plane DM interaction $D_{\parallel}$. In Fig.~\ref{Hall_honey}(a) the magnons are topological by setting $D_z=0.1$ meV. By coupling to the phonons the magnitude of $\kappa_{xy}$ increases. We ascribe this  to the redistribution of Berry curvature among the bands as shown in Fig.~\ref{Berry1}. In particular, the contribution of the Berry curvature of magnons (peaks at $\mbox{K}_{\pm}$) in $\kappa_{xy}$ is washed out by positive values of Berry curvatures (peaks appearing between $\Gamma-\mathbf{K_{\pm}}$) near the band hybridization. However, new sharp features appearing between $\Gamma-M$ (see inset in Fig.~\ref{honey:effective}) belonging to states at different energies give rise to the increase of the thermal Hall conductivity. On the other hand, the hybridization leads to finite thermal Hall response even when both magnons and phonons are trivial. The results are shown in Fig.~\ref{Hall_honey}(b), where we set $D_z=0$ and sublattice potential $J\delta S=0.1$ meV endowing the magnons with trivial band topology. Upon hybridization, the lowest energy bands becomes topologically nontrivial. As explained in the discussion of Fig.~\ref{Berry2}, the lower band acquires negative Berry curvatures throughout the BZ. The emergent Berry curvature underlies the finite values of $\kappa_{xy}$ shown in Fig.~\ref{Hall_honey}(b). By increasing $D_{\parallel}$ from $0.2$ to $1.0$ meV, the magnitude of Hall response increases. Stronger hybridization gives rise to larger values of thermal Hall conductivity. Indeed, for the trivial bands, the in-plane DM interaction act like gauge fields for phonons, which upon the time-reversal symmetry breaking yields topological bands. We use the same parameters in both figures. One, however, observes that the thermal Hall conductivity of a posterior topological magnon is an order of magnitude larger than that of trivial modes.  

Note that the obtained values of thermal Hall conductivity $\kappa_{xy}\simeq 10^{-12}$ WK$^{-1}$ is 
consistent with values obtained in Refs.[\onlinecite{Zhang:PRL2019,Park:Nano2020}]. For that the heat current density $J^{Q}_{xy}$ should be understood as surface current density in 2D systems with unit Watt/Length. However, if the latter is to be measured as bulk current density with unit Watt/Area, our obtained values should be divided by the thickness of the samples. For layers with thickness of order of $t\sim 10^{-10}-10^{-9}$m, the thermal Hall conductivity becomes $\kappa_{xy}\simeq 10^{-2}-10^{-3}$ WK$^{-1}$m$^{-1}$.

\section{Hybrid magnon-phonon model: the Kagome lattice\label{Sec:Kagome}}

\begin{figure*}[!htb]
  \centering
  \includegraphics[width=\linewidth]{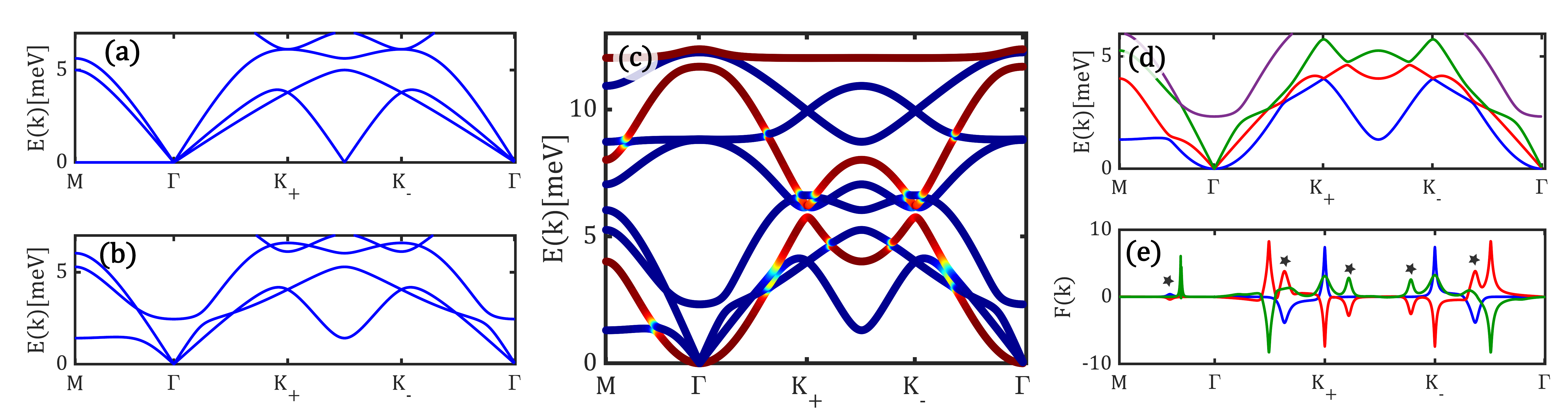}
  \caption{(a) Low-energy phonon spectrum of regular Kagome lattice with only nearest-neighbor inter-ion potential, $\hbar\Omega=5$ meV and $\hbar\Omega'=0$. The zero modes along $\Gamma-M$ are the floppy modes. (b) The same band structure including the second-neighbor inter-ion potential, $\hbar\Omega'=1$ meV. The floppy modes become dispersive. (c) The magnetoelastic spectrum of hybrid magnon (the reddest color) and phonon (the bluest color) modes. The avoided crossings are visible by change in the color. We show the lowest magnetoelastic bands with different colors, blue, red, green, and purple in panel (d). The corresponding Berry curvatures of three lowest bands are shown in panel (e) using the same colors as in (d). In (e) the numbers on vertical axis should be understood as $\mathrm{sgn}(F(\mathbf{k}))\ln(1+|F(\mathbf{k})|)$, and the asterisks marked the locations of avoided crossings. }
    \label{rKagome}
\end{figure*}

In this section we present hybrid magnon-phonon modes in the kagome lattice. A piece of this lattice is shown in Fig.~\ref{lattice}. Each unit cells contain three sites yielding a three magnon bands and six phonon bands considering only the vibration within the plane. The kagome lattice with coordination number $z=4$ is at the isostatic point separating floppy from rigid behaviors\cite{pnas:sun2012}. The zero-energy floppy modes become dispersive by adding second-neighbor inter-ionic potentials, i. e., by increasing the coordination number. Alternatively, since the lattice is at the isostatic point, the unit cells are twisted and the floppy modes becomes dispersive and the rigidity condition arises \cite{pnas:sun2012}. We call the former case as a regular Kagome lattice and the latter one as a twisted one. Below, we study both cases when coupled to magnons.           

\subsection{Regular Kagome Lattice}

The regular kagome lattice lattice is shown in Fig.~\ref{lattice}. We follow the procedures outlined in Sec.~\ref{ModelH} to obtain the magnetoelastic Hamiltonian on the kagome lattice. The details of different parts of the Hamiltonian are relegated to Appendix \ref{rKagome2}. Unlike the honeycomb lattice, on the kagome lattice the nearest-neighbor out-of-plane DM interaction $D_z$ is allowed. In Fig.~\ref{rKagome}(a) we show the lowest energy bands of the phonon spectrum by considering only the nearest-neighbor inter-ion potentials $V(\mathbf{R}_{ij})$ in \eqref{V_ion}. As seen, there are zero floppy modes along the $\Gamma-M$ direction and all symmetry- related directions in the BZ implying that the lattice is at the isostatic point. By adding second-neighbor inter-ion potentials, the floppy modes become dispersive with typical energy of the order of $\hbar\Omega'$ as shown in Fig.~\ref{rKagome}(b), and therefore only the zero modes at $\Gamma$ associated with the trivial translations and rotations are retained. 

The magnon spectrum consists of three energy bands whose their band topology are characterized by Chern numbers $c=\pm1, 0$ in the presence of DM interaction $D_z$. As in the honeycomb lattice the in-plane DM interaction $D_{\parallel}$ hybridizes the magnons and phonons as shown in Fig.~\ref{rKagome}(c). The two lowest magnon bands cross the phonon bands at multiple points, as illustrated by light colors, where the avoided crossings and magnetopolaron states are formed. It is also seen that the dispersive modes along $\Gamma-M$ direction, otherwise being zero-energy modes in the absence of coupling to magnons, get hybridized with magnons. The lowest magnetoelastic modes are shown in Fig.~\ref{rKagome}(d) along with the corresponding Berry curvature in Fig.~\ref{rKagome}(e). We used the same colors, blue, red, and green for three lowest bands and their Berry curvatures. It is clearly seen that states near the avoided crossings lead to large enhancement in the Berry curvature. The peaks associated with avoided crossings are marked by asterisks. Note that the zero crossings between phonon bands, like those at $\mbox{K}_{\pm}$, also result in large Berry curvature, e.g., the blue and red peaks at $\mbox{K}_{\pm}$. However, since the the bands are degenerate at these crossings, they don't contribute in thermal Hall conductivity, to be discussed later on. The green peaks at $\mbox{K}_{\pm}$ are associated with the topological magnon states, while the adjacent green peaks appearing along $\Gamma-\mbox{K}_{+}$, $\mbox{K}_{+}-\mbox{K}_{-}$, and $\mbox{K}_{-}-\Gamma$ result from the avoided crossings. Since the sates around these points are split in energy, they would influence the thermal Hall response as we demonstrated for the case of honeycomb lattice.

\subsection{Twisted Kagome Lattice}

\begin{figure*}[!htb]
  \centering
  \includegraphics[width=\linewidth]{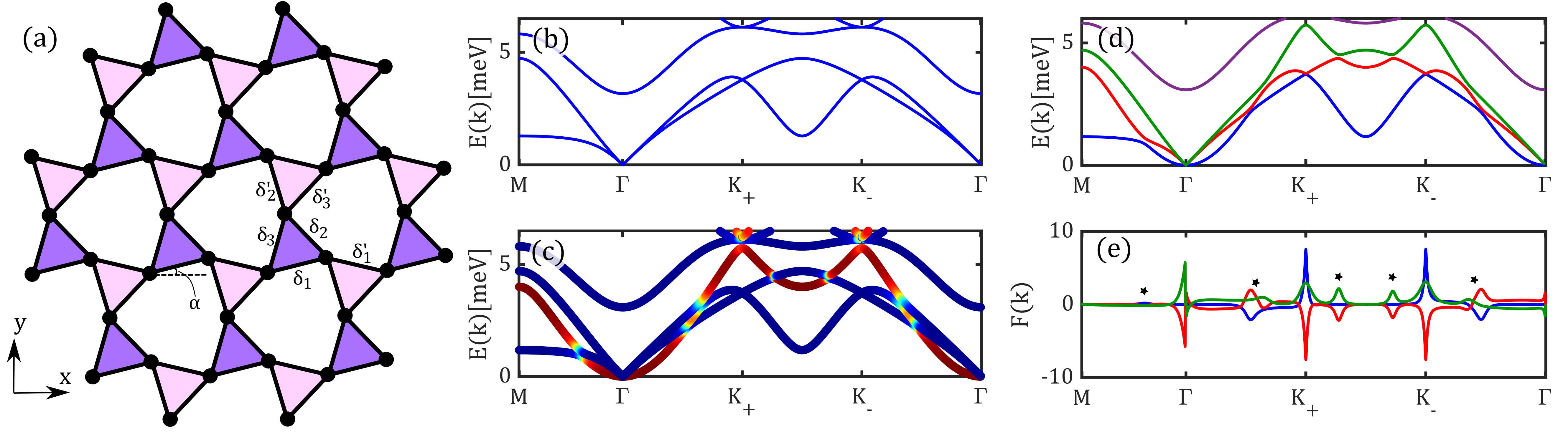}
  \caption{(a) A piece of twisted Kagome lattice. The up (purple) and down (pink) triangles are twisted by $\alpha>0$ and $-\alpha$, respectively. The unit cell still contains three sites. (b) The low-energy phonon spectrum of the twisted Kagome lattice with $\hbar\Omega=5$ meV and $\hbar\Omega'=0$. The twist removes the floppy modes otherwise having zero energy between $\Gamma-M$ direction. (c) The magnetoelastic spectrum of twisted lattice. (d) same as the spectrum in (c) but the bands are colored as blue, red, and green for the three lowest ones, and the corresponding Berry curvatures are shown in (e). The asterisks show the regions of avoided crossing where the Berry curvatures are created for magnetoelastic bands with different energies. In (e) the numbers on vertical axis should be understood as $\mathrm{sgn}(F(\mathbf{k}))\ln(1+|F(\mathbf{k})|)$.}
    \label{tKagome}
\end{figure*}

As discussed in preceding section the floppy zero modes become dispersive by including second-neighbor potential with frequency $\Omega'>0$. In this case the lattice retains its $C_6$ symmetry. In the absence of second-neighbor potential, i.e., $\Omega'=0$, the floppy zero modes can also be removed by twisting the unit cells as shown in Fig.~\ref{tKagome}(a), where up and down triangles are twisted by an angle $\alpha>0$ and $-\alpha$, respectively. The symmetry reduces to $C_3$ and the lattice vectors are squeezed to $|\mathbf{a}_{1,2}|=2a\cos\alpha$ where $a$ is the bond length. We obtain the phonon Hamiltonian as before, and the detailes are given in Appendix \ref{tKagome2}. The phonon energy spectrum is shown in Fig.~\ref{tKagome}(b) for $\alpha=\pi/12$ and clearly shows that the floppy zero modes between $\Gamma-M$ are replaced by finite dispersive modes\cite{pnas:sun2012}, hence yielding a stable lattice. This phonon spectrum should be compared with the one shown in Fig.~\ref{rKagome}(b), where the floppy zero modes are removed by taking $\Omega'>0$. Therefore, either adding second-neighbor interaction or twisting the cells disperses the floppy modes.

Next, we couple phonons to the magnons. The low-energy part of the hybrid spectrum is shown in Fig.~\ref{tKagome}(c). The details of the magnon-phonon coupling matrices are given in Appendix \ref{tKagome2}. At the crossings the bands are split off due to magnon-phonon coupling similar to those in the regular Kagome lattice. To study the Berry curvature of energy bands we replot the low-energy bands in Fig.~\ref{tKagome}(d) and used the blue, red, and green colors to label the lowest to highest ones. The corresponding Berry curvature along the same high-symmetry lines is shown in Fig.~\ref{tKagome}(e). Besides the features associated with the bands crossings of the phonon spectrum, we note that the avoided crossings marked by asterisks lead to creation of Berry curvatures with opposite signs. The sharp features around $\mbox{K}_{\pm}$ are due to phonon band degeneracies, and as we pointed out they don't contribute to the thermal Hall response. The features around $\Gamma$ point is due to the nearly degenerate  phonon bands. Other peaks emanating from the magnon-phonon band hybridization are split in energy, hence having different contribution in the thermal Hall conductivity. Therefore, the redistribution of the Berry curvatures from the purely topological magnon bands to the magnetoelastic bands could influence the thermal Hall measurements as we discuss in the next subsection. 

\begin{figure}[b]
   \center
    \includegraphics[width=\linewidth]{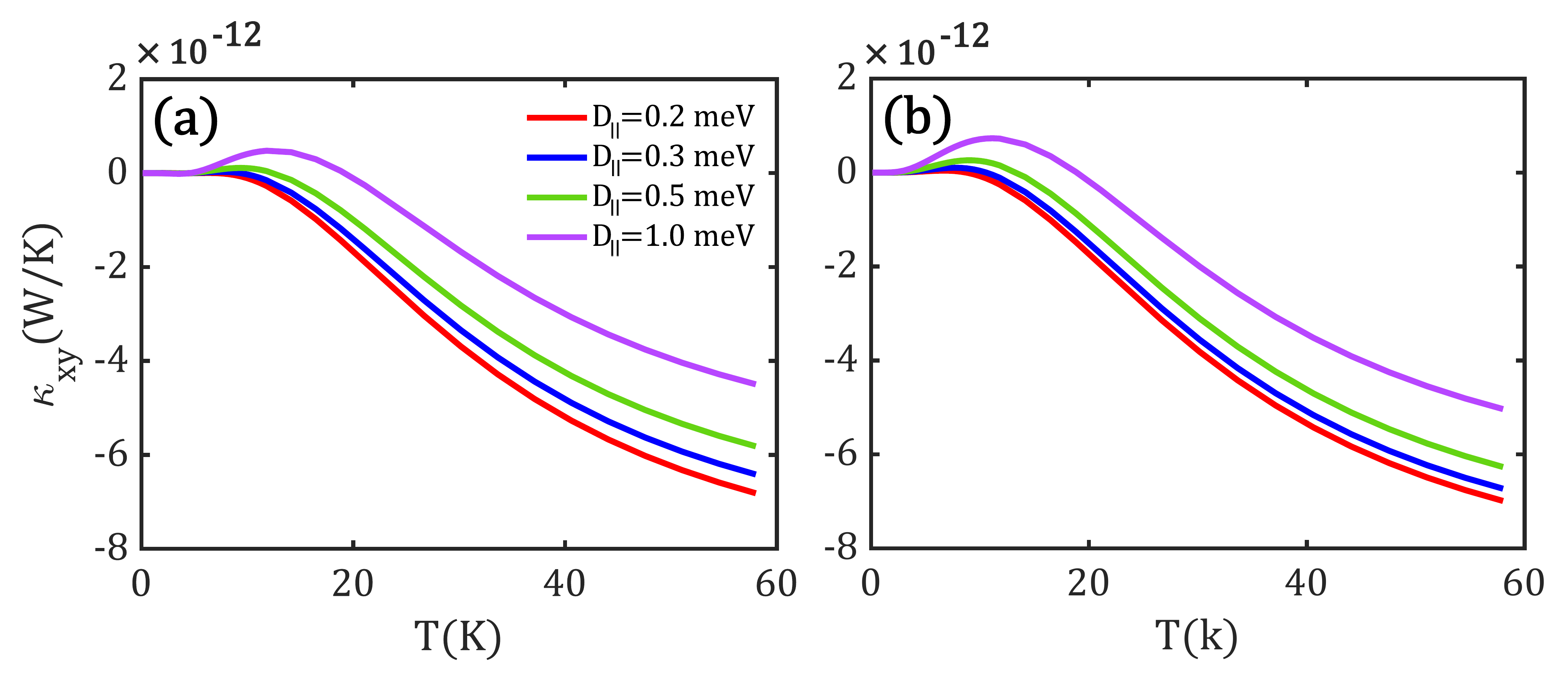}
    \caption{ The temperature dependence of thermal Hall conductivity $\kappa_{xy}$ of (a) regular and (b) twisted kagome lattices. In both cases we set $D_z=0.1$ meV, hence topological magnons, and colored curves from bottom to top correspond to different values of $D_{\parallel}$: 0.2 (red), 0.3 (blue), 0.5 (green), 1.0 (purple) meV. For the twisted lattice we set $\alpha=\pi/12$.} \label{Hall_kagome}
\end{figure}

\subsection{Thermal Hall Conductivity: the Kagome lattices}
Informed by the magnetoelastic bands and Berry curvatures, we show the thermal Hall conductivity of the regular and the twisted kagome lattices in Fig.~\ref{Hall_kagome}(a) and Fig.~\ref{Hall_kagome}(b), respectively. Both lattices show the similar behaviors. For temperatures below $\sim$10 K the response is nearly zero especially for small values of in-plane DM interaction. By increasing the temperature the response acquires large values of order of $\sim 10^{-11}$WK$^{-1}$. It also shows that the in-plane DM interaction $D_{\parallel}$ adds more positive contributions to $\kappa_{xy}$, hence the magnitude of the response decreases especially at higher temperatures. This is attributed to the thermal populations of higher bands with negative Berry curvatures, e.g., the red plots in Fig.~\ref{rKagome}(e) and  Fig.~\ref{tKagome}(e) near the band hybridizations. We also observe that for large values of $D_{\parallel}$, the thermal Hall response changes sign by temperatures. We note that a sign change in $\kappa_{xy}$ has been observed in the kagome magnet Cu(1,3-benzenedicarboxylate) \cite{Ong:PRL2015}. Note that this occurs in the presence of magnon-phonon couplings, and thus, any interpretation of possible observation of sign change in the measurements should take both carriers into account. Usually such sign change is related to the energy distribution of states near peaks of the Berry curvatures. While the total Berry curvature of a band could be positive, e.g., the lowest energy band of magnons on the kagome lattice\cite{Mook:prb2014}, the coupling to phonons may create states at low energy with negative Berry curvatures and and states with positive Berry curvature at higher energy. Therefore, the occupations of states by increasing the temperature leads to the sign change in thermal Hall response.

\section{Conclusions\label{conclusion}}
This work is partly motivated by recent observation of topological bosonic modes in several compounds such as  Lu$_2$V$_2$O$_7$ \cite{Onose:Science2010}, planar kagome magnets Cu(1,3-benzenedicarboxylate) \cite{Ong:PRL2015} and YMn$_6$Sn$_6$\cite{Zhang:prb2020}, Tb$_3$Ga$_5$O$_{12}$\cite{Strohm:PRL2005,Inyushkin:JETP2007}, Tb$_2$Ti$_2$O$_7$\cite{Hirschberger:Science2015}, and CrCl$_3$ \cite{Pocs:PRR2020} as discussed in the introduction. The heat conduction supplemented to the system by a temperature gradiant however excites both magnons and phonons. Therefore, it is interesting to investigate the interplay of such collective modes and the issue of magnon-phonon couplings in determining the thermal properties of the system. 

We theoretically introduced a hybrid magnon-phonon models on the honeycomb and kagome lattices, for which the lattice structures allow for topological magnons to arise by adding phase windings, resulting from out-of-plane DM interaction, to the propagating magnons. The in-plane components of DM interaction arising from the mirror symmetry breaking, however, couple magnons and phonons and generate magnetoelastic modes. Our effective description of magnon-phonon hybridization clearly demonstrates that the change of wave function components from magnons to phonons and vice versa leads to a pronounced enhancement of the Berry curvature near the avoided crossings. The observation is that the Berry curvatures, while appearing with opposite signs, belong to states with different energies. We found that this latter point and the magnon-phonon induced redistribution of Berry curvature among the energy bands have significant effects on the thermal Hall conductivity. For the honeycomb lattice the magnon-phonon coupling increases the thermal Hall response. In particular, we found that for topological magnon bands, the hybridization to phonons increases the response, an order of magnitude larger than the trivial bands. For both regular and twisted kagome lattices we found that the magnon-phonon coupling gives rise to a sign change of the thermal Hall response at low temperature. Therefore, we anticipate that in any interpretations of responses both magnons and phonons should be treated on equal footings.

\section{ Acknowledgments}
The authors would like to acknowledge the support from Sharif University of Technology under Grant No. G960208 and Iran National Elite Federation.

\newpage
\appendix
\begin{widetext}
\section{First and second neighbor potential matrices between the sites of the honeycomb lattice\label{Vmatrix}}
For the first neighbors the matrices are,

\begin{align}
V_{10}=\begin{pmatrix}
  3/2 & 0 & 0 & 0\\
  0 & 3/2 & 0 & -1\\
  0 & 0 & 3/2 & 0\\
  0 & -1 & 0 & 3/2  
 \end{pmatrix},~~ V_{11}=\begin{pmatrix}
  0 & 0 & -3/2 & -\sqrt{3}/2\\
  0 & 0 & -\sqrt{3}/2 & -1/2\\
  0 & 0 & 0 & 0\\
  0 & 0 & 0 & 0  
 \end{pmatrix}, ~~~ V_{12}=\begin{pmatrix}
  0 & 0 & -3/2 & \sqrt{3}/2\\
  0 & 0 & \sqrt{3}/2 & -1/2\\
  0 & 0 & 0 & 0\\
  0 & 0 & 0 & 0  
 \end{pmatrix},
\end{align}  
and for the second neighbors we obtain $V_{20}=3\mathbf{1}_{4\times4}$ and remaining matrices are as 

\begin{align}
 V_{21}=\begin{pmatrix}
  -1/2 & -\sqrt{3}/2 & 0 & 0\\
  -\sqrt{3}/2 & -3/2 & 0 & 0\\
  0 & 0 & -1/2 & -\sqrt{3}/2\\
  0 & 0 & -\sqrt{3}/2 & -1/2 
 \end{pmatrix}, ~~~ V_{22}=\begin{pmatrix}
  -1/2 & \sqrt{3}/2 & 0 & 0\\
  \sqrt{3}/2 & -3/2 & 0 & 0\\
  0 & 0 & -1/2 & \sqrt{3}/2\\
  0 & 0 & \sqrt{3}/2 & -1/2 
 \end{pmatrix},~~V_{23}=\begin{pmatrix}
  -2 & 0 & 0 & 0\\
  0 & 0 & 0 & 0\\
  0 & 0 & -2 & 0\\
  0 & 0 & 0 & 0 
 \end{pmatrix}.
\end{align}  

\section{The magnetoelastic Hamiltonian of the regular Kagome lattice \label{rKagome2}}
The regular lattice is shown in Fig.~\ref{lattice}. Assuming a ferromagnetic classical ground state and within the linear spin-wave theory, the magnon Hamiltonian reads as  

\begin{align}
H_M = \frac{1}{2}\sum_{\mathbf{ k}} \tilde{\psi}^{\dagger} _{\mathbf{ k}} \tilde{H}_M(\mathbf{ k})\tilde{\psi} _{\mathbf{ k}},
\end{align}
 where, using $A$, $B$, and $C$ to label three sites of the unit cell, $\tilde{\psi} _{\mathbf{ k}} =(b_{\mathbf{k}A}, b_{\mathbf{k}B}, b_{\mathbf{k}C}, b^{\dagger}_{-\mathbf{k}A}, b^{\dagger}_{-\mathbf{k}B}, b^{\dagger}_{-\mathbf{k}C})^{t}$ and 

\begin{align}\label{app:HM}
 \tilde H_{M}(\mathbf{k})=\begin{pmatrix}
  Q(\mathbf{k}) & 0_{3\times3}\\
 0_{3\times3} & Q^{t}(-\mathbf{k}) 
 \end{pmatrix},~~~Q(\mathbf{k})=\begin{pmatrix}
  f_1 & f_{12}(\mathbf{k})&f_{13}(\mathbf{k})\\
 f_{12}^*(\mathbf{k}) & f_{1}&f_{23}(\mathbf{k})\\
f_{13}^*(\mathbf{k}) & f_{23}^*(\mathbf{k})&f_{1}
 \end{pmatrix},
\end{align}
with

\begin{align}
f_1=4JS,~f_{12}=-J_ne^{i\phi}\left(1+e^{-i\mathbf{k}\cdot\mathbf{a}_1} \right),~f_{13}=-J_ne^{-i\phi}\left(1+e^{-i\mathbf{k}\cdot\mathbf{a}_2} \right),~f_{23}=-J_ne^{i\phi}\left(1+e^{-i\mathbf{k}\cdot(\mathbf{a}_2-\mathbf{a}_1)} \right),
\end{align}
and $\phi=\tan^{-1}(D/J)$. The phonon Hamiltonian reads as
\begin{align}
H_{Ph}=\frac{1}{2}\sum_{\mathbf{k}}\phi^t_{-\mathbf{k}} H_{Ph}(\mathbf{k})\phi_{\mathbf{k}},
\end{align}
where $\phi _{\mathbf{ k}}=\left(u_{\mathbf{k}A}^x,u_{\mathbf{k}A}^y,u_{\mathbf{k}B}^x,u_{\mathbf{k}B}^y,u_{\mathbf{k}C}^x,u_{\mathbf{k}C}^y, p_{-\mathbf{k}A}^x,p_{-\mathbf{k}A}^y,p_{-\mathbf{k}B}^x,p_{-\mathbf{k}B}^y,p_{-\mathbf{k}C}^x,p_{-\mathbf{k}C}^y \right)^t $ and 
\begin{align}
H_{ph}(\mathbf{k})=\hbar\Omega\begin{pmatrix}
  V(\mathbf{k})& 0_{6\times6}\\
 0_{6\times6} & \mathbf{1}_{6\times6}
 \end{pmatrix}.
\end{align}

Here, $V(\mathbf{k})=[V_{nn}(\mathbf{k})+\xi^2V_{nnn}(\mathbf{k})]/2$ describes the nearest and next-nearest neighbor inter-ion potentials:

\begin{align}
&V_{nn}(\mathbf{k})=V_{0}(\mathbf{k})+V_{1}(\mathbf{k})+V_{2}(\mathbf{k})+V_{3}(\mathbf{k}),\\
&V_{nnn}(\mathbf{k})=V'_{0}(\mathbf{k})+V'_{1}(\mathbf{k})+V'_{2}(\mathbf{k})+V'_{3}(\mathbf{k}),
\end{align}

\begin{align}
&V_{0}(\mathbf{k})=2V_{0},~V_{1}(\mathbf{k})=V_{1}e^{-i\mathbf{k}\cdot\mathbf{a}_1}+V^{\dagger}_{1}e^{i\mathbf{k}\cdot\mathbf{a}_1},\\
&V_2(\mathbf{k})=V_2e^{-i\mathbf{k}\cdot\mathbf{a}_2}+V^{\dagger}_2e^{i\mathbf{k}\cdot\mathbf{a}_2},~V_3(\mathbf{k})=V_3e^{-i\mathbf{k}\cdot(\mathbf{a}_1-\mathbf{a}_2)}+V^{\dagger}_3e^{i\mathbf{k}\cdot(\mathbf{a}_1-\mathbf{a}_2)},
\end{align}

\begin{align}
&V'_{0}(\mathbf{k})=2V'_{0},~V'_{1}(\mathbf{k})=V'_{1}e^{-i\mathbf{k}\cdot\mathbf{a}_1}+V'^{\dagger}_{1}e^{i\mathbf{k}\cdot\mathbf{a}_1},\\
&V'_2(\mathbf{k})=V'_2e^{-i\mathbf{k}\cdot\mathbf{a}_2}+V'^{\dagger}_2e^{i\mathbf{k}\cdot\mathbf{a}_2},~V'_3(\mathbf{k})=V'_3e^{-i\mathbf{k}\cdot(\mathbf{a}_1-\mathbf{a}_2)}+V'^{\dagger}_3e^{i\mathbf{k}\cdot(\mathbf{a}_1-\mathbf{a}_2)}.
\end{align}

The nearest-neighbor matrices are:
\begin{align}
&V_0 = \frac{1}{4}\begin{pmatrix}
10&2\sqrt{3}&- 4&0&- 1&- \sqrt{3}\\
2\sqrt{3}&6&0&0&-\sqrt{3}&- 3\\
- 4&0&10&- 2\sqrt{3}&- 1&\sqrt{3}\\
0&0&- 2\sqrt{3}&6&\sqrt{3}&- 3\\
-1&- \sqrt{3}&- 1&\sqrt{3}&\sqrt{3}&0\\
-\sqrt{3}&-3&\sqrt{3}&-3&0&12
\end{pmatrix},~V_1 = \begin{pmatrix}
0&0&- 2&0&0&0\\
0&0&0&0&0&0\\
0&0&0&0&0&0\\
0&0&0&0&0&0\\
0&0&0&0&0&0\\
0&0&0&0&0&0
\end{pmatrix},\\
&V_2 = \frac{1}{4}\begin{pmatrix}
0&0&0&0&-2&-2\sqrt{3}\\
0&0&0&0&-2\sqrt{3}&-6\\
0&0&0&0&0&0\\
0&0&0&0&0&0\\
0&0&0&0&0&0\\
0&0&0&0&0&0
\end{pmatrix},~V_3 =\frac{1}{4} \begin{pmatrix}
0&0&0&0&0&0\\
0&0&0&0&0&0\\
0&0&0&0&-2&2\sqrt{3}\\
0&0&0&0&2\sqrt{3}&- 6\\
0&0&0&0&0&0\\
0&0&0&0&0&0
\end{pmatrix}.
\end{align}

The next-nearest-neighbor matrices are:
\begin{align}
&V'_0 = \frac{1}{4}\begin{pmatrix}
6&- 2\sqrt{3}&0&0&0&0\\
-2\sqrt{3}&{10}&0&0&0&0\\
0&0&6&2\sqrt{3}&0&0\\
0&0&2\sqrt{3}&{10}&0&0\\
0&0&0&0&{12}&0\\
0&0&0&0&0&4
\end{pmatrix},~
V'_1 = \frac{1}{4}\begin{pmatrix}
0&0&0&0&- 6&2\sqrt{3}\\
0&0&0&0&2\sqrt{3}&-2\\
0&0&0&0&0&0\\
0&0&0&0&0&0\\
0&0&-6&- 2\sqrt{3}&0&0\\
0&0&-2\sqrt{3}&-2&0&0
\end{pmatrix},\\
&V'_2 = \frac{1}{4}\begin{pmatrix}
0&0&0&0&0&0\\
0&0&0&-8&0&0\\
0&0&0&0&-6&-2\sqrt{3}\\
0&0&0&0&-2\sqrt{3}&-2\\
0&0&0&0&0&0\\
0&0&0&0&0&0
\end{pmatrix},~
V'_3 = \frac{1}{4}\begin{pmatrix}
0&0&0&0&{ - 6}&{2\sqrt 3 }\\
0&0&0&0&{2\sqrt 3 }&{ - 2}\\
0&0&0&0&0&0\\
0&{ - 8}&0&0&0&0\\
0&0&0&0&0&0\\
0&0&0&0&0&0
\end{pmatrix}.
\end{align}

The magnon-phonon coupling Hamiltonian $H_c$ is:
\begin{align}
H_c=\sum_{\mathbf{k}}\phi^{\dagger}_{ph}(\mathbf{k}) H_{c}(\mathbf{k})\phi_{m}(\mathbf{k}),
\end{align}
where $\phi_{ph}(\mathbf{k})=(u^x_{\mathbf{k}A}, u^y_{\mathbf{k}A}, u^x_{\mathbf{k}B}, u^y_{\mathbf{k}B},u^x_{\mathbf{k}C}, u^y_{\mathbf{k}C})^{t}$, $\phi_{m}(\mathbf{k})=(\delta S^x_{\mathbf{k}A}, \delta S^y_{-\mathbf{k}A}, \delta S^x_{\mathbf{k}B}, \delta S^y_{-\mathbf{k}B}, \delta S^x_{\mathbf{k}C}, \delta S^y_{-\mathbf{k}C})^{t}$, and

\begin{align}
H_{c}(\mathbf{k})=\begin{pmatrix}
2(T_1+T_3)& -T_1\left(1+e^{-i\mathbf{k}\cdot\mathbf{a}_1} \right)&-T_3\left(1+e^{-i\mathbf{k}\cdot\mathbf{a}_2} \right)\\
 -T_1\left(1+e^{i\mathbf{k}\cdot\mathbf{a}_1} \right)& 2(T_2+T_1)& -T_2\left(1+e^{-i\mathbf{k}\cdot(\mathbf{a}_2-\mathbf{a}_1)}\right)\\
-T_3\left(1+e^{i\mathbf{k}\cdot\mathbf{a}_2} \right)& -T_2\left(1+e^{i\mathbf{k}\cdot(\mathbf{a}_2-\mathbf{a}_1)}\right)& 2(T_3+T_2)
\end{pmatrix}.
\end{align}

The above $T$-matrices describing the magnon-phonon couplings along $\boldsymbol{\delta}_1=\hat{x}$, $\boldsymbol{\delta}_2=\hat{x}/2+\sqrt{3}/2\hat{y}$, and $\boldsymbol{\delta}_3=-\hat{x}/2+\sqrt{3}/2\hat{y}$ connecting the nearest-neighbor sites, are 

\begin{align}
T_1=\zeta D_{\parallel}S\begin{pmatrix}
-\gamma&0\\
0&1
\end{pmatrix}, T_2=\frac{\zeta SD_{\parallel}}{4}\begin{pmatrix}
  3-\gamma & \sqrt{3}(1+\gamma) \\
  \sqrt{3}(1+\gamma) & 1-3\gamma  
 \end{pmatrix}, T_3=\frac{\zeta SD_{\parallel}}{4}\begin{pmatrix}
  3-\gamma & -\sqrt{3}(1+\gamma) \\
  -\sqrt{3}(1+\gamma) & 1-3\gamma  
 \end{pmatrix}.
\end{align}

\section{The magnetoelastic Hamiltonian of the twisted Kagome lattice \label{tKagome2}}

For the twisted Kagome lattice the magnetic Hamiltonian is as \eqref{app:HM}. For the phonon part the nearest-neighbor $V$-matrices are as follows:

\begin{align}\nonumber
V_0 = \frac{\cos^2\alpha}{4}\begin{pmatrix}
10&2\sqrt{3}&-4&0&-1&-\sqrt 3\\
2\sqrt{3}&6&0&0&-\sqrt 3&-3\\
-4&0&10&-2\sqrt{3}&-1&\sqrt{3}\\
0&0&-2\sqrt{3}&6&\sqrt{3}&-3\\
-1&-\sqrt{3}&-1&\sqrt{3}&4&0\\
-\sqrt{3}&-3&\sqrt{3}&-3&0&12
\end{pmatrix}+\frac{\sin^2\alpha}{4}\begin{pmatrix}
6&-2\sqrt{3}&0&0&-3&\sqrt 3\\
-2\sqrt{3}&10&0&-4&\sqrt 3&-1\\
0&0&6&2\sqrt{3}&-3&-\sqrt{3}\\
0&-4&2\sqrt{3}&10&-\sqrt{3}&-1\\
-3&\sqrt{3}&-3&-\sqrt{3}&12&0\\
\sqrt{3}&-1&-\sqrt{3}&-1&0&4
\end{pmatrix}\\
+\frac{\sin2\alpha}{4}\begin{pmatrix}
0&0&0&-2&\sqrt{3}& 1\\
0&6&-2&0&1&-\sqrt{3}\\
0&-2&0&0&-\sqrt{3}&1\\
-2&0&0&0&1&\sqrt{3}\\
\sqrt{3}&1&-\sqrt{3}&1&0&0\\
1&-\sqrt{3}&1&\sqrt{3}&0&0
\end{pmatrix},
\end{align}

\begin{align}
V_1 = \cos^2\alpha\begin{pmatrix}
0&0&-2&0&0&0\\
0&0&0&0&0&0\\
0&0&0&0&0&0\\
0&0&0&0&0&0\\
0&0&0&0&0&0\\
0&0&0&0&0&0
\end{pmatrix}+\sin^2\alpha\begin{pmatrix}
0&0&0&0&0&0\\
0&0&0&-2&0&0\\
0&0&0&0&0&0\\
0&0&0&0&0&0\\
0&0&0&0&0&0\\
0&0&0&0&0&0
\end{pmatrix}
+\sin2\alpha\begin{pmatrix}
0&0&0&1&0&0\\
0&0&1&0&0&0\\
0&0&0&0&0&0\\
0&0&0&0&0&0\\
0&0&0&0&0&0\\
0&0&0&0&0&0
\end{pmatrix},
\end{align}

\begin{align}
V_2 = \frac{\cos^2\alpha}{4}\begin{pmatrix}
0&0&0&0&-2&-2\sqrt 3\\
0&0&0&0&-2\sqrt 3&-6\\
0&0&0&0&0&0\\
0&0&0&0&0&0\\
0&0&0&0&0&0\\
0&0&0&0&0&0
\end{pmatrix}+\frac{\sin^2\alpha}{4}\begin{pmatrix}
0&0&0&0&-6&2\sqrt 3\\
0&0&0&0&2\sqrt 3&-2\\
0&0&0&0&0&0\\
0&0&0&0&0&0\\
0&0&0&0&0&0\\
0&0&0&0&0&0
\end{pmatrix}
+\frac{\sin2\alpha}{4}\begin{pmatrix}
0&0&0&0&-2\sqrt{3}&-2\\
0&0&0&0&-2&2\sqrt{3}\\
0&0&0&0&0&0\\
0&0&0&0&0&0\\
0&0&0&0&0&0\\
0&0&0&0&0&0
\end{pmatrix},
\end{align}

\begin{align}
V_3 = \frac{\cos^2\alpha}{4}\begin{pmatrix}
0&0&0&0&0&0\\
0&0&0&0&0&0\\
0&0&0&0&-2&2\sqrt{3}\\
0&0&0&0&2\sqrt{3}&-6\\
0&0&0&0&0&0\\
0&0&0&0&0&0
\end{pmatrix}+\frac{\sin^2\alpha}{4}\begin{pmatrix}
0&0&0&0&0&0\\
0&0&0&0&0&0\\
0&0&0&0&-6&-2\sqrt{3}\\
0&0&0&0&-2\sqrt{3}&-2\\
0&0&0&0&0&0\\
0&0&0&0&0&0
\end{pmatrix}
+\frac{\sin2\alpha}{4}\begin{pmatrix}
0&0&0&0&0&0\\
0&0&0&0&0&0\\
0&0&0&0&2\sqrt{3}&-2\\
0&0&0&0&-2&-2\sqrt{3}\\
0&0&0&0&0&0\\
0&0&0&0&0&0
\end{pmatrix}.
\end{align}

For the magnon-phonon coupling, there are $T$-matrices along nearest-neighbor bonds $\boldsymbol{\delta}_{1,2,3}$ and $\boldsymbol{\delta}'_{1,2,3}$ shown in Fig.~\ref{tKagome}(a). We denote the corresponding matrices by $T_{1,2,3}$ and $T'_{1,2,3}$, respectively, and they read as follows:

\begin{align}
T_1 = \zeta D_{\parallel}S\begin{pmatrix}
1-(1+\gamma)\cos^2\alpha&-(1+\gamma)\sin\alpha\cos\alpha\\
-(1+\gamma)\sin\alpha\cos\alpha&1-(1+\gamma)\sin^2\alpha
\end{pmatrix},
\end{align}

\begin{align}
T_2 = \frac{\zeta D_{\parallel}S}{4}\begin{pmatrix}
4-(1+\gamma)(1+2\sin^2\alpha+\sqrt{3}\sin2\alpha)&(1+\gamma)(\sqrt{3}\cos2\alpha+\sin2\alpha )\\
(1+\gamma)(\sqrt{3}\cos2\alpha+\sin2\alpha )&4-(1+\gamma)(1+2\cos^2\alpha-\sqrt{3}\sin2\alpha)
\end{pmatrix},
\end{align}

\begin{align}
T_3 = \frac{\zeta D_{\parallel}S}{4}\begin{pmatrix}
4-(1+\gamma)(1+2\sin^2\alpha-\sqrt{3}\sin2\alpha)&-(1+\gamma)(\sqrt{3}\cos2\alpha-\sin2\alpha )\\
-(1+\gamma)(\sqrt{3}\cos2\alpha-\sin2\alpha )&4-(1+\gamma)(1+2\cos^2\alpha+\sqrt{3}\sin2\alpha)
\end{pmatrix},
\end{align}

\begin{align}
T'_1 = \zeta D_{\parallel}S\begin{pmatrix}
1-(1+\gamma)\cos^2\alpha&(1+\gamma)\sin\alpha\cos\alpha\\
(1+\gamma)\sin\alpha\cos\alpha&1-(1+\gamma)\sin^2\alpha
\end{pmatrix},
\end{align}

\begin{align}
T'_2 = \frac{\zeta D_{\parallel}S}{4}\begin{pmatrix}
4-(1+\gamma)(1+2\sin^2\alpha-\sqrt{3}\sin2\alpha)&-(1+\gamma)(\sqrt{3}\cos2\alpha+\sin2\alpha )\\
-(1+\gamma)(\sqrt{3}\cos2\alpha+\sin2\alpha )&4-(1+\gamma)(1+2\cos^2\alpha+\sqrt{3}\sin2\alpha)
\end{pmatrix},
\end{align}

\begin{align}
T'_3 = \frac{\zeta D_{\parallel}S}{4}\begin{pmatrix}
4-(1+\gamma)(1+2\sin^2\alpha+\sqrt{3}\sin2\alpha)&-(1+\gamma)(\sqrt{3}\cos2\alpha+\sin2\alpha )\\
-(1+\gamma)(\sqrt{3}\cos2\alpha+\sin2\alpha )&4-(1+\gamma)(1+2\cos^2\alpha-\sqrt{3}\sin2\alpha)
\end{pmatrix}.
\end{align}

And the magnon-phonon coupling Hamiltonian is: 

\begin{align}
H_{c}(\mathbf{k})=\begin{pmatrix}
(T_1+T_3+T'_1+T'_3)& -\left(T_1+T'_1e^{-i\mathbf{k}\cdot\mathbf{a}_1} \right)&-\left(T_3+T'_3e^{-i\mathbf{k}\cdot\mathbf{a}_2} \right)\\
 -\left(T_1+T'_1e^{i\mathbf{k}\cdot\mathbf{a}_1} \right)& (T_2+T_1+T'_2+T'_1)& -\left(T_2+T'_2e^{-i\mathbf{k}\cdot(\mathbf{a}_2-\mathbf{a}_1)}\right)\\
-\left(T_3+T'_3e^{i\mathbf{k}\cdot\mathbf{a}_2} \right)& -\left(T_2+T'_2e^{i\mathbf{k}\cdot(\mathbf{a}_2-\mathbf{a}_1)}\right)& (T_3+T_2+T'_3+T'_2)
\end{pmatrix}.
\end{align}

\end{widetext}

\newpage
%

\end{document}